\documentclass[prb,twocolumn,showpacs,preprintnumbers,amsmath,amssymb,superscriptaddress]{revtex4-1}
\usepackage{graphicx}
\usepackage{dcolumn}
\usepackage{bm}
\usepackage{subfigure}
\usepackage{color}

\newcommand{\comment}[1]{}
\newcommand{\COMMENT}[1]{}

\def\v{{\bf v}}

\newcommand{\BFCA}{Ba(Fe$_{1-x}$Co$_x$)$_2$As$_2$\ }

\newcommand{\BFAP}{BaFe$_2$(As$_{1-x}$P$_x$)$_2$\ }

\graphicspath{{./figures/}}

\begin{document}

\title{Volovik effect in a highly anisotropic multiband superconductor: experiment and theory}
\author{Y. Wang}
\author{J.S. Kim}
\author{G. R. Stewart}
\author{P.J. Hirschfeld}
\affiliation{Department of Physics, University of Florida, Gainesville, Florida 32611, USA}
\author{S. Graser}
\affiliation{Center for Electronic Correlations and Magnetism, Institute of Physics,\\
University of Augsburg, D-86135 Augsburg, Germany}
\author{S. Kasahara}
\author{T. Terashima}
\affiliation{Research Center for Low Temperature and Materials Sciences, Kyoto University, Kyoto 606-8501,
Japan}
\author{Y. Matsuda}
\author{T. Shibauchi}
\affiliation{Department of Physics, Kyoto University, Sakyo-ku, Kyoto 606-8502, Japan}
\author{I. Vekhter}
\affiliation{Department of Physics and Astronomy, Louisiana State University, Baton Rouge, Louisiana 70803-4001, USA}
\date{\today}
\begin{abstract}
We present measurements of the specific heat coefficient $\gamma$($\equiv C/T)$ in the low temperature limit
as a function of an applied magnetic field for the Fe-based superconductor BaFe$_2$(As$_{0.7}$P$_{0.3}$)$_2$.
We find both a linear regime at higher fields and a limiting square root $H$ behavior at very low fields. The
crossover from a Volovik-like $\sqrt{H}$ to a linear field dependence can be understood from a multiband
calculation in the quasiclassical approximation assuming gaps with different momentum dependence on the hole-
and electron-like Fermi surface sheets.
\end{abstract}

\maketitle

\section{Introduction}
The symmetry and detailed structure of the gap function in the recently discovered iron
pnictide~\cite{kamihara08} and chalcogenide~\cite{Hsu08} high temperature superconductors is still under
discussion.  Across an increasingly numerous set of materials families, as well as within each family where
superconductivity can be tuned by doping or pressure, experimental indications are that there is no universal
gap structure.~\cite{Wen11,[{}][{; to appear in Rev. Mod. Phys. (2011).}]stewart11} Instead, the
superconducting gap appears to be remarkably sensitive to details of the normal state properties. This
``intrinsic sensitivity"~\cite{Kemper10} may be due to the unusual Fermi surface topology, consisting of
small hole and electron pockets, and to the probable $A_{1g}$ symmetry of the superconducting gap which
allows a continuous deformation of the order parameter structure from a fully gapped system to one with nodes
(for a review see, e.g. Ref.~\onlinecite{HKMreview}). It is important to keep in mind, though, that another
possibility to account for the observed variability is that different experiments on the same material may
probe selectively different Fermi surface regions and hence different gaps within the system.

The Ba-122 family of materials has been intensively studied because large high quality single crystals are
relatively easy to produce.~\cite{stewart11,Kasahara10}  Within this family, the isovalently substituted
system \BFAP with a maximum $T_c$ of 31 K  is particularly intriguing because it exhibits a phase diagram and
transport properties remarkably similar to the heterovalently doped system \BFCA and displays many signatures
of apparent quantum critical behavior at optimal doping.~\cite{Kasahara10,jiang09,Shishido10} In the
superconducting state, penetration depth,~\cite{Hashimoto10} NMR spin-lattice relaxation,~\cite{Nakai10}
thermal conductivity temperature dependence,~\cite{Hashimoto10} and thermal conductivity angular field
variation~\cite{Yamashita11} show clear indications of nodal behavior. Surprisingly, a linear field
dependence of the specific heat Sommerfeld coefficient $\gamma$ was measured\cite{Kim10} on optimally doped
samples from the same batch. Such a behavior is expected for a fully gapped single band superconductor since
the fermionic excitations from the normal cores of vortices provide the only contribution to $\gamma$ at low
$T$, and the number of these vortices scales linearly with the field $H$.  It was argued in
Ref.~\onlinecite{Kim10} that the specific heat measurement might be consistent with the other experiments
suggesting nodes if the heavy hole sheets in the material were fully gapped, while the gaps on the lighter
electron sheets were nodal. In such a case the $\gamma\sim\sqrt{H}$ behavior would be difficult to observe in
experiment.

In this paper, we report new experimental data on the magnetic field dependence of the specific heat of
optimally doped \BFAP samples.  We have extended our previous measurements to 15~T to higher fields up to 35
T ($\approx\frac{2}{3}H_{c2}(0)$), where we find a continuation of the linear behavior reported earlier.
However, more precise measurements at low fields have revealed the presence of a Volovik-like $\sqrt{H}$ term
which persists roughly over a range of 4 T, crossing over to a linear behavior above this scale.~\footnote{In
contrast to BaFe$_2$(As$_{1-x}$P$_x$)$_2$, recent high field measurements on underdoped ($x=0.045$) and
overdoped ($x=0.103$) BaFe$_{2-x}$Co$_x$As$_2$ have found that the specific heat coefficient varies
approximately as $H^{0.7}$ all the way up to $H_{c2}(0)$. J. S. Kim, G. R. Stewart, K. Gofryk, F. Ronning,
and A. S. Sefat, to be published.} The observation of this term, consistent with nodes in the superconducting
gap, therefore supports claims made in earlier work,\cite{Hashimoto10,Nakai10,Yamashita11} without the need
to assume an extremely large mass on the hole pockets.

Theoretical estimates using the Doppler shift method for isotropic gaps given in Ref.~\onlinecite{Bang10}
were oversimplified, but did show the need for a more thorough analysis of anisotropic multiband systems, and
stimulated further experimental work, both of which we report here.  The theoretical difficulties can be seen
easily by considering a simple two-band model with two distinct gaps $\Delta_1$ and $\Delta_2$, where we
assume for the moment that $\Delta_2>\Delta_1$.  If the two bands are uncoupled, the two gaps correspond to
two independent coherence lengths $\xi_i\simeq v_{F,i}/(\pi \Delta_i)$, where $i=1,2$, and two independent
``upper critical fields" $H_{c2,i}$.  Vortex core states of the large gap $\Delta_2$ are confined to cores of
radius $\sim\xi_2$.  For fields in the range $H_{c2,1} \lesssim H \lesssim H_{c2,2}$, the vortex cores of the
small gap will overlap, while the large gap cores will still be well separated. Note that if $\Delta_1$ is
very small (these considerations also describe crudely nodal gaps), this field range can be wide and extend
to quite low fields. On the other hand, methods of studying quasiparticle properties in superconductors are
typically adapted to calculating near $H_{c1}$ or $H_{c2}$, i.e. in the limit of widely separated or nearly
overlapping vortices. The current problem apparently contains elements of both situations. In the absence of
interband coupling, of course, one can use different methods, corresponding to the appropriate field regimes,
for the distinct bands. For coupled Fermi surfaces, however, such an approach is not viable. In the immediate
vicinity of the transition, where the Ginzburg-Landau expansion is valid, there is a single length scale
controlling the vortex structure.~\cite{Kogan:2011} At low temperatures, where the measurements are carried
out, however, the distinct length scales likely survive, although they are modified by the strength of the
interband coupling, see below. Possible anisotropy of the gap on one or more Fermi surface sheets complicates
the picture even further. We show here that judicious use of the quasiclassical approximation even with
simplifying assumptions about the vortex structure can provide a general framework for the description of
this problem, and a semiquantitative understanding of the new data on the \BFAP system.  We compare our
results with those obtained by Doppler-shift methods, and show that if properly implemented this method also
gives reasonable qualitative results in the low field range.

This joint theory-experiment paper is organized as follows.  We first present our experimental results on the
\BFAP system in Section~\ref{sec:expt}  and compare to our previous results, as well as to data by other
groups on the related heterovalently doped Ba-122 materials. In Section~\ref{sec:model}, we discuss the
two-band quasiclassical model we use to study the system, and in Section~\ref{sec:results} we give our
results. Finally in Section~\ref{sec:conclusions} we present our conclusions.

\section{Experimental}\label{sec:expt}
Tiny platelet crystals of BaFe$_2$(As$_{0.7}$P$_{0.3}$)$_2$ were prepared as described in
Ref.~\onlinecite{Kasahara10}. Subsequent measurements on crystals extracted from various positions in the
crucible using x-ray diffraction and energy dispersion (EDX) analysis give a phosphorous concentration of
$32.9\pm0.4\%$. A further test of the homogeneity of the crystals from a given growth batch is the
measurement of the susceptibility at the superconducting transition as shown in Fig.~1 of
Ref.~\onlinecite{Kim10}. Here, for a collage of $\sim 150$ mg of these crystals a rather narrow transition
was observed. A collage of $18$ mg of these microcrystals was then mounted on a sapphire disk using GE7031
varnish. Approximately $75\%$ of the crystals had the magnetic field perpendicular to the a-b plane (the
plane of the crystals), whereas the remaining crystals were randomly oriented.  The sapphire disk was mounted
in our time constant method calorimeter,~\cite{stewart83,*Andraka89} and the specific heat from $0.4$ to
$7$~K in fields from $0$ to $35$~T was measured. Additionally, the specific heat of a standard (high purity
Au) was measured in fields up to $14$~T. Results on the standard (not shown) indicate agreement with
published values to within $\pm 3\%$ in all fields.

\subsection{Results and Discussion}
\begin{figure}[t]
\includegraphics[width=0.45\textwidth,trim=4mm 6mm 4mm 2mm]{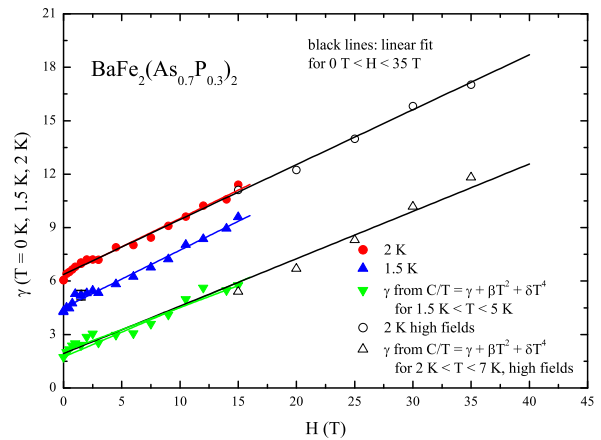}
\caption{\label{fig:P_dopedGamma35T} (Color online)  The original specific heat data\cite{Kim10} on
BaFe$_2$(As$_{0.7}$P$_{0.3}$)$_2$ as a function of field up to 15~T (solid symbols) with data from present
work between 15~T and 35~T (open symbols). Note the agreement between the linear, $C/T\propto H$,
extrapolation of the 15~T (colored lines~\cite{Kim10}) and 35~T (black lines, present work) results. We
extract $\gamma$ from the data using two (equivalent) methods: (a)  by making an extrapolation
$C/T=\gamma+\beta T^2+\delta T^4$ from 2~K and above or (b) by taking the value of $C/T$ at 1.5 and 2~K. The
temperature restriction eliminates both the influence of the anomaly and the field-induced nuclear
contribution (see text), negligible for $H\leq4$~T above 1~K. The absolute accuracy of these data is $\pm
5\%$ while the precision of the data is approximately $\pm2\%$. In addition, additional data with finer
gradations in the measured fields up to $4$~T were taken to explore the low field non-linear behavior. These
data are shown on an expanded scale in Fig.~\ref{fig:P_dopedGammaLoField}. }
\end{figure}
\begin{figure}
\includegraphics[width=0.45\textwidth,trim=4mm 6mm 4mm 2mm]{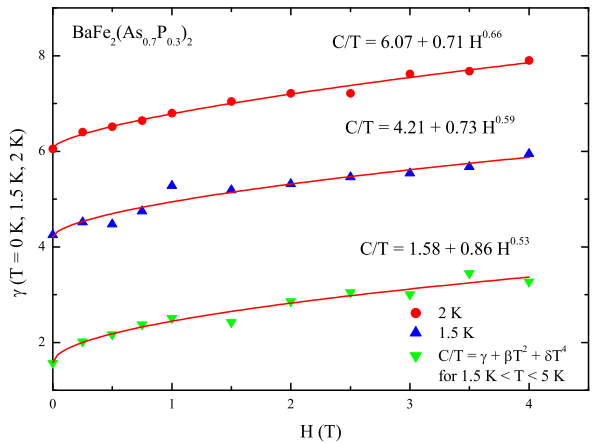}
\caption{\label{fig:P_dopedGammaLoField} (Color online)  Low field $\gamma$ data up to 4~T from
Fig.~\ref{fig:P_dopedGamma35T} on an expanded scale for T=2~K (blue), 1.75~K (red) and 1.5~K (black symbols).
Green symbols are asymptotic $\lim_{T\to 0} C/T$ determined over the range 1.5~K$<T<$5~K. The fitting
functions of the data are labeled beside the curves. Best power law fits to field dependence are shown in
each case. }
\end{figure}

The specific heat coefficient $\gamma\equiv C/T$ of BaFe$_2$(As$_{0.7}$P$_{0.3}$)$_2$ for $0\leq H\leq 35$~T
is shown by the open triangles in Fig.~\ref{fig:P_dopedGamma35T}. There is a small low temperature anomaly in
the specific heat data below about 1.4~K (discussed in detail in \onlinecite{Kim10}). Such anomalies have
been observed in other FePn samples,~\cite{Kim09} and in some cases, e.g. in BaFe$_{2-x}$Co$_x$As$_2$, they
show a rather strong magnetic field dependence.~\cite{Kim09} However, as discussed in our previous
report~\cite{Kim10} of the data up to 15T, the anomaly in BaFe$_2$(As$_{0.7}$P$_{0.3}$)$_2$ is approximately
field independent. Note that the small anomaly in the specific heat appears to vanish above 1.4~K, i.e. does
not affect the estimate for $\gamma$ shown in Figs.~\ref{fig:P_dopedGamma35T}
and~\ref{fig:P_dopedGammaLoField} using data from 1.5~K and above.

In order to have a closer look at the low field dependence of the specific heat, these data are shown on an
expanded scale in Fig.~\ref{fig:P_dopedGammaLoField}.  In our analysis below, we focus on the asymptotic
$T\rightarrow 0$ behavior since it is directly related to the density of states at the Fermi level, which is
easy to calculate reliably, and since it gives essentially the same field dependence as the nonzero $T$ data.

\section{Model}\label{sec:model}

\subsection{Quasiclassical approximation}
The quasiclassical (Eilenberger) approximation~\cite{eilenb68,larkin69,serene83} is a powerful tool to
describe the electronic properties of the superconducting state on the scales large compared to the lattice
spacing, provided the condition $k_F\xi\gg 1$ is satisfied. Here $k_F$ is the Fermi momentum and $\xi$ the
coherence length. Since in this limit we can think of quasiparticles as propagating coherently along a
well-defined trajectory in real space, this method is particularly well suited to address the inhomogeneous
situations, such as the vortex state of type-II superconductors (SCs). An alternative and frequently used
approach to the vortex state is to take into account the (classical) shift of the quasiparticle energy due to
the local supercurrent flow.  Such an approximation, often referred to as the Doppler-shift approach, is
valid for nodal SCs with considerable weight of extended quasiparticle excitations ouside the vortex cores.
Using this method, Volovik showed that for superconductors with line nodes these extended quasiparticle
excitations lead to a non-linear magnetic field dependence of the spatially averaged residual density of
states $N(\omega=0,H)\propto N_0\sqrt{H/H_{c2}}$, the result known as the Volovik effect.~\cite{volovik93}
This behavior was first confirmed by measurements of the specific heat~\cite{KAMoler:1997,YWang:2001} and by
subsequent calculations within the quasiclassical approximation for both a single vortex in a $d$-wave
SC~\cite{Ichioka96,Schopohl95} and for a vortex lattice.~\cite{Ichioka97,Ichioka99}  Both quasiclassical and
Doppler-shift methods fail at the lowest temperatures due to quantum effects~\cite{FranzTesanovic}, but in
known systems with $T_c\ll E_F$ these effects are negligible in practice.  Both methods have successfully
explained at a semiquantitative level the magnetic field dependence of the specific heat and thermal
conductivity in a wide variety of unconventional superconductors.~\cite{Vekhter08} It was also shown that the
accurately calculated quasiparticle excitation spectrum is consistent with STM studies of the electronic
structure around a vortex core.~\cite{Ichioka97}

Many experimental techniques which are sensitive to the low-energy density of states, such as thermal
conductivity, specific heat, and NMR relaxation rate, can be used to draw conclusions about the possible
existence and the momentum dependence of quasiparticle excitation in the bulk of iron-based superconductors
(FeSCs) and thus about the structure of the superconducting gap and the distribution of gap nodes. The low
$T$ limit of the Sommerfeld coefficient in an applied magnetic field, $\gamma(H)$, is directly proportional
to the spatially averaged local density of states (LDOS) at the Fermi level. The Doppler-shift method has
been used to calculate the LDOS for a two-band SC with two isotropic gaps of unequal size
$\Delta_S\neq\Delta_L$ and to give an interpretation of the  experimental data available at that
time.\cite{Bang10} However, the Doppler-shift approach cannot account properly for the contributions from the
states in the vortex core that have a very large weight in the net DOS and hence gives a quantitatively and
sometimes qualitatively inaccurate description of the electronic structure of the vortex. For example, in a
simple $d$-wave superconductor the spatial tails of the low-energy density of states around the vortex are
aligned in the wrong directions.\cite{Dahm02}  To obtain a quantitative fit to the specific heat data
presented in the previous section and to allow for a more decisive conclusion about the gap structure of
BaFe$_2$(As$_{0.7}$P$_{0.3}$)$_2$, we will therefore use the quasiclassical approximation, which we will
briefly review in the following paragraphs.

In the quasiclassical method, the Gorkov Green's functions are integrated with respect to the quasiparticle
energy measured from the Fermi level. The normal and anomalous components $g(\mathbf{r},\theta,i\omega_n)$
and $f(\mathbf{r},\theta,i\omega_n)$ of the resulting propagator $\hat{g}$ obey the coupled Eilenberger
equations
\begin{subequations}\label{eq:Eilenberger}
\begin{align}
  &\left[ 2\left( i \omega_n+\frac{e}{c}\mathbf{v}_F\cdot\mathbf{A(r)} \right)+i\hbar\mathbf{v}_F\cdot\nabla
  \right]f(\mathbf{r},\theta,i\omega_n)\notag\\
    &\quad=2ig(\mathbf{r},\theta,i\omega_n)\Delta(\mathbf{r},\theta),\\
  &\left[ 2\left( i \omega_n+\frac{e}{c}\mathbf{v}_F\cdot\mathbf{A(r)} \right)-i\hbar\mathbf{v}_F\cdot\nabla
  \right]\bar{f}(\mathbf{r},\theta,i\omega_n)\notag\\
    &\quad=2ig(\mathbf{r},\theta,i\omega_n)\Delta^{*}(\mathbf{r},\theta),
\end{align}
\end{subequations}
that have to be complemented by the normalization condition
\begin{align}\label{eq:normCond}
  \hat{g}^2\equiv\begin{pmatrix}g &f  \\ \bar{f} &-g\end{pmatrix}^2= \hat{1}\, .
\end{align}
Here $\Delta(\mathbf{r},\theta)$ is the order parameter, $\mathbf{A(r)}$ the vector potential, $\mathbf{v}_F$
is the Fermi velocity at the location at the Fermi surface labeled by $\theta$, and $\omega_n=(2n+1)\pi k_B
T$ are the fermionic Matsubara frequencies. For two-dimensional cylindrical Fermi surfaces such as considered
below, $\mathbf{v}_F=v_F\hat{\mathbf{k}}$ where $\hat{\mathbf{k}}=(\cos\theta,\sin\theta)$ and $\theta$ is
the angle measured from the [100] direction. In that case it is natural to write the position vector in
cylindrical coordinates, $\mathbf{r}=(\rho,\phi,z)$, where $\phi$ is the winding angle around the vortex in
real space.

Making use of the symmetries~\cite{schopohl98} of the quasiclassical propagator \footnote{Note that our
notation of $g$, $f$, and $\bar{f}$ differs from the one used in Ref.~\onlinecite{schopohl98}. Under the
transformation $g\to -i\pi g$, $f\to \pi f$, and $\bar{f}\to -\pi \bar{f}$ the notation in
Ref.~\onlinecite{schopohl98} passes into our notation.}
\begin{subequations}
\begin{align}
  \bar{f}(\mathbf{r},\mathbf{k}_F,i\omega_n)&=f^{*}(\mathbf{r},\mathbf{k}_F,-i\omega_n),\\
  f(\mathbf{r},-\mathbf{k}_F,-i\omega_n)&=f(\mathbf{r},\mathbf{k}_F,i\omega_n),\\
  g(\mathbf{r},\mathbf{k}_F,i\omega_n)&=g^{*}(\mathbf{r},\mathbf{k}_F,-i\omega_n),
\end{align}
\end{subequations}
the diagonal part of the normalization condition (\ref{eq:normCond}) can be written in a more explicit form
as
\begin{align}
  [g(\mathbf{r},\theta,i\omega_n)]^2+f(\mathbf{r},\theta,i\omega_n)f^*(\mathbf{r},\theta+\pi,i\omega_n)=1\, .
\end{align}

Instead of solving the complicated coupled Eilenberger equations everywhere in space, we follow
Refs.~\onlinecite{Schopohl95,schopohl98} and parameterize the quasiclassical propagator along real space
trajectories $\mathbf{r}(x)=\mathbf{r}_0+x\hat{\mathbf{v}}_F$ by a set of scalar amplitudes $a(x)$ and
$b(x)$,
\begin{align}
  \hat{g}(\mathbf{r}(x))=\frac{1}{1+a(x)b(x)}\begin{pmatrix}
    1-a(x)b(x)  &2a(x)\\
    2b(x)     &-1+a(x)b(x)
  \end{pmatrix}.
\end{align}
These amplitudes obey numerically stable Riccati equations,
\begin{subequations}\label{eq:Riccati}
\begin{align}
  v_F\partial_x a(x)+[2\tilde{\omega}_n+\Delta^*(x)a(x)]a(x)-\Delta(x) &=0,\\
  v_F\partial_x b(x)-[2\tilde{\omega}_n+\Delta(x)b(x)]b(x)+\Delta^*(x) &=0\,.
\end{align}
\end{subequations}
For the single vortex problem the spatial dependence vanishes far away from the vortex core, and hence we
have the initial conditions
\begin{subequations}
\begin{align}
  a(-\infty)&=\frac{\Delta(-\infty)}{\omega_n+\sqrt{\omega_n^2+|\Delta(-\infty)|^2}}\, ,\\
  b(+\infty)&=\frac{\Delta^*(+\infty)}{\omega_n+\sqrt{\omega_n^2+|\Delta(+\infty)|^2}}\, .
\end{align}
\end{subequations}
Here we have set $\hbar=1$ and we have introduced the modified Matsubara frequencies
$i\tilde{\omega}_n(x)=i\omega_n+(e/c)\mathbf{v}_F\cdot\mathbf{A}(x)$. Since the modification of the Matsubara
frequencies due to the external field is of the order of $1/\kappa^2$ where $\kappa = \lambda_L/\xi$ is the
ratio of the London penetration depth and the coherence length the term proportional to $\mathbf{A}(x)$ in
Eq. (\ref{eq:Riccati}) can be neglected for strong type-II superconductors.

After an analytic continuation of the Matsubara frequencies to the real axis, $i \omega_n \to \omega +
i\delta$, the local density of states can be calculated as the Fermi surface average of the quasiclassical
propagator
\begin{align}\label{eq:DOS}
  N(\mathbf{r})=N_0\int_0^{2\pi}\frac{d\theta}{2\pi}\text{Re}
  \left(\frac{1-ab}{1+ab}\right)_{i\omega_n\to \omega+i\delta},
\end{align}
where $N_0$ is the normal density of states at the Fermi energy. To obtain stable numerical solutions we use a small
imaginary part $\delta=0.02T_c$ in the analytical continuation, where $T_c$ is the critical temperature
of the superconductor.

\subsection{Two-band model}
The Fermi surface of the optimally doped BaFe$_2$(As$_{0.7}$P$_{0.3}$)$_2$ consists of multiple Fermi surface
sheets. DFT calculations showed that there are three concentric hole cylinders in the center of the Brillouin
zone ($\Gamma$ point) and two electron pockets at the zone corner ($\mathrm{X}$ point).\cite{Kuroki2011}
Laser ARPES measurements~\cite{Shimojima11} found a superconducting order parameter that is fully gapped with
comparably sized gaps on each hole pocket of the order of $\Delta_h/k_BT_c\sim 1.7$.  Taking into account the
results from thermal conductivity~\cite{Hashimoto10,Yamashita11} and NMR measurements~\cite{Nakai10} as well
as the measurements of the specific heat coefficient in low fields presented above, that all consistently
report evidence for low-energy quasiparticles, this ARPES result implies a nodal gap on the electron pockets.

For numerical convenience we adopt below a two-band model, distinguishing only between electron and hole
pockets. Inclusion of all Fermi surface sheets then only enters as a weighting factor for the electron and
hole pocket contributions as we discuss in the following section. We take the gaps on the electron and hole
pockets in the form $\Delta_{1,2}(\theta)=\Delta_0^{e,h}\Phi_{1,2}(\theta)$, where the angle $\theta$
parameterizes the appropriate Fermi surface, assumed to be cylindrical. We assume an anisotropic gap on the
electron pocket\cite{Mishra09} $\Phi_1(\theta)=(1+r\cos2\theta)/\sqrt{1+r^2/2}$, and an isotropic gap around
the hole Fermi surface, $\Phi_2(\theta)=1$. If the anisotropy factor $r>1$, the superconducting gap in the
electron band, $\Delta_1(\theta)$,  has accidental nodes; if $r=0$, $\Delta_1(\theta)$ is isotropic like
$\Delta_2(\theta)$.

First we assume $\Delta_0^e=\Delta_0^h$, as is often found by ARPES (at this writing there are no ARPES
results on this material which resolve the gap on the electron pocket). Since we consider well separated
electron and hole bands, we can solve the Riccati equations, Eqs.~(\ref{eq:Riccati}),  for the two
propagators separately, and the only coupling of the pockets is via the self-consistency equations on the
order parameter, see below. With this in mind we normalize the energy and length for the electron and hole
bands by the gap amplitudes $\Delta_0^{e}$ and $\Delta_0^{h}$, and the coherence lengths
$\xi_0^e=v_F^e/\Delta_0^e$ and $\xi_0^h=v_F^h/\Delta_0^h$ respectively. Fermi velocities therefore appear as
an input. DFT calculations for a comparable Ba-122 system~\cite{Mishra2011} give $v_F^h=1.979\times 10^5$ m/s
and $v_F^e=3.023\times 10^5$ m/s, i.e. $v_F^h/v_F^e=\xi_0^h/\xi_0^e=0.65$. In our analysis we keep this ratio
but reduce the value of both Fermi velocities by a factor of 5 to approximately account for the mass
renormalization of this system near optimal doping.\cite{Shishido10,Yoshida11}  This reduction also  gives a
roughly  correct value of the $c$-axis upper critical field $H_{c2}\sim 50$~T. In the limit of negligible
coupling between the bands, the upper critical field is determined by the overlap of the vortices with
smallest core size,
\begin{align}\label{eq:RHdepend}
  \frac{R}{\min\{\xi_0^e,\xi_0^h\}}=\frac{R}{\xi_0^h}=\sqrt{\frac{H_{c2}}{H}}.
\end{align}
Below we solve the Eilenberger equations and determine the density of states for an isolated vortex and for
each band separately. In a two-band system the spatial profile of the quasiparticle states on the electron
and hole bands is controlled by the respective coherence lengths, and therefore spatial averaging weighs the
contributions of the bands differently compared to the DOS of a system with a single or two equal coherence
lengths. This is the most significant difference compared to a single-band model.

\begin{figure}
\includegraphics[width=0.42\textwidth]{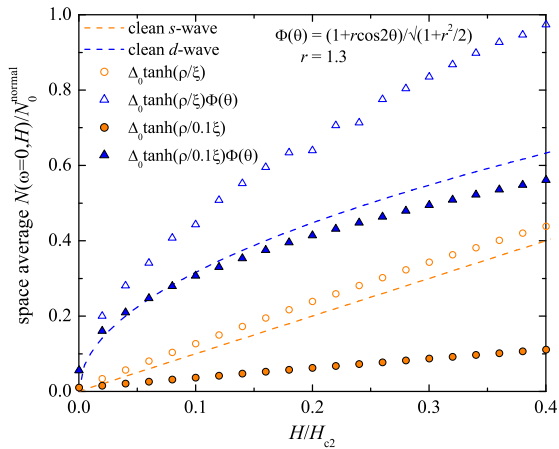}
\caption{\label{fig:onebandXi}(Color online) The spatially averaged ZDOS, normalized to the normal state
value $N(\omega=0)/N_0$ for a nodeless (orange) and a nodal (blue) single-band superconductor. The dashed
lines show the idealized linear $H$ and $\sqrt{H}$ behavior for a clean $s$-wave and $d$-wave SC,
respectively. The symbols are numerical results for a single band SC with an isotropic $s$-wave gap (circles)
and a strongly anisotropic nodal gap (triangles). Additionally we compare results with (solid symbols) and
without (open symbols) taking into account the vortex core reduction due to the Kramer-Pesch effect. Here we
have ignored the field dependence of the superconducting gap, i.e. $\Delta(H)=\Delta_0$.}
\end{figure}

The superconducting order parameters in the two bands are related by the interband component of the pairing
interaction. We consider a general coupling matrix in the factorized form,
$\lambda_{\nu\mu}(\theta,\theta')=\lambda_{\nu\mu} \Phi_\nu(\theta)\Phi_\mu(\theta')$, where $\mu, \nu=1,2$
and $\lambda_{\nu\mu}\equiv V_{\nu\mu}N_\mu$. Here $V_{11}=V_{e}$ and $V_{22}=V_{h}$ are the intraband
pairing interactions in the electron and the hole band, respectively, while $V_{12}=V_{eh}$ is the interband
interaction. $N_\mu$ is the normal density of states at the Fermi level. Then the gap equation for an
inhomogeneous superconductor is
\begin{align}\label{eq:gapEqn}
  \Delta_\nu(\mathbf{r})=2\pi T\sum_{\mu=1,2}\lambda_{\nu\mu} \sum_{\omega_n>0}^{\omega_c}
    \langle \Phi_\mu(\theta)f_\mu(\mathbf{r},\theta,i\omega_n) \rangle_\theta\, .
\end{align}
Here $\Delta_\nu(\mathbf{r})$ is the momentum independent part of the gap function;
$\Delta_{1,2}=\Delta_0^{e,h}$ at $T=0$ and $H=0$.

In the vortex state the self consistent determination of the spatially dependent order parameter is a complex
task. Since we are interested in relatively low fields, when the vortices are well separated,  we solve the
Eilenberger equations for the order parameter that is assumed to have a single vortex form,
\begin{subequations}\label{eq:gapforms}
\begin{align}\label{eq:gapforms1}
  \Delta^{e}(\vec{\rho},H;\theta)&=\Delta_1(H)\tanh\left(\frac{\rho}{0.1\xi_0^e}\right)
      \frac{1+r\cos2\theta}{\sqrt{1+r^2/2}},\\\label{eq:gapforms2}
  \Delta^{h}(\vec{\rho},H)&=\Delta_2(H)\tanh\left(\frac{\rho}{0.1\xi_0^h}\right)\,.
\end{align}
\end{subequations}
Here $\vec{\rho}=(\rho,\phi)$ is the two-dimensional projection of the radius vector in cylindrical
coordinates, and factor of 0.1 is introduced to approximate the shrinking of the core size in the
self-consistent treatment at low temperatures (Kramer-Pesch effect~\cite{KramerPesch1,KramerPesch2}). This
single vortex ansatz provides a qualitatively correct description of the low-field state, close to what is
found by full numerical solution.~\cite{Dahm02} To account for the suppression of the bulk order parameter by
the magnetic field, we determine the coefficients $\Delta_{1,2}(H)$ from the Pesch
approximation,~\cite{Pesch_approx1} where in the presence of an Abrikosov lattice the diagonal components of
the Green's function by its value averaged over a unit cell of the vortex lattice. This approximation proven
to give reliable results over a considerable range of magnetic fields and is incorporated into our approach.

Note that our ansatz for the order parameter becomes quantitatively inaccurate for strong interband coupling
in the regime of applicability of the Ginzburg-Landau theory since the core sizes of the two bands approach
each other.~\cite{Zhitomirsky:2004} We verified in a fully self-consistent calculation that in the parameter
range that we use the corresponding effect on the specific heat is of order 1\% or less and hence can be
neglected. We therefore use Eq.~(\ref{eq:gapforms}) hereafter.

To proceed we substitute Eq. (\ref{eq:gapforms}) into Eq. (\ref{eq:Riccati}), solve for $a(x)$ and $b(x)$,
and use Eq. (\ref{eq:DOS}) to find the local density of states $N(\vec{\rho},H)$. To approximate the specific
heat coefficient, we evaluate the spatial average of the zero energy local density of states
\begin{align}\label{eq:dosHInt}
  \bar{N}(H)=\int_0^{2\pi}d\phi \int_0^{R}d\rho \, \rho \frac{N(\vec{\rho},H)}{\pi R^2 N_0},
\end{align}
where the intervortex distance $R$ depends on $H$ as described by Eq. (\ref{eq:RHdepend}). The total density
of states is then given as
\begin{align}
  \bar{N}(H)_\text{tot}&=\frac{w_e\bar{N}^e(H)+w_h\bar{N}^h(H)}{w_e+w_h} \label{eq:sumDOSeh}
\end{align}
where $w_e/w_h=2N_0^e/N_0^h=2\zeta$ if we consider, for example, two electron Fermi surface sheets in the folded
Brillouin zone and denote $\zeta\equiv N_0^e/N_0^h=v_F^h/v_F^e=0.65=\lambda_{21}/\lambda_{12}$.

\section{Results}\label{sec:results}
\begin{figure*}
\centering
  \subfigure{\includegraphics[trim=0 0 12mm 0,height=50mm]{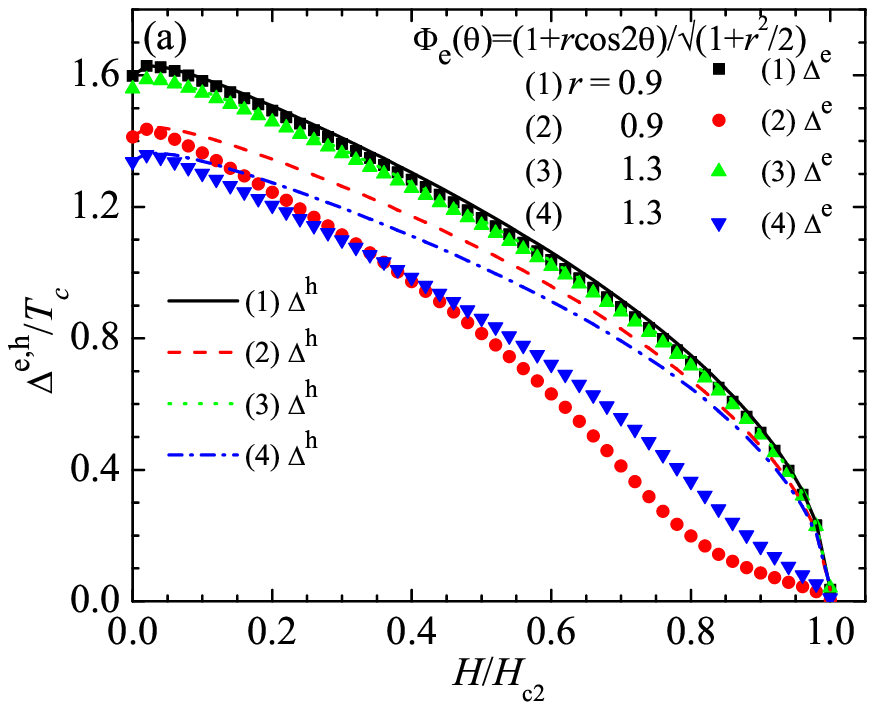}}
  \subfigure{\includegraphics[height=47.5mm]{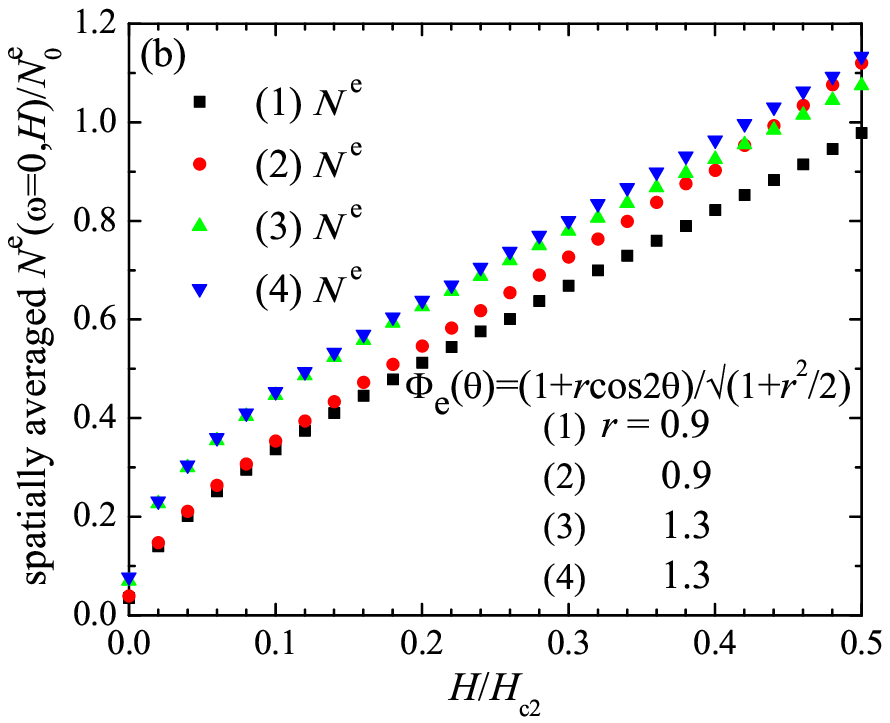}}
  \subfigure{\includegraphics[height=47.5mm]{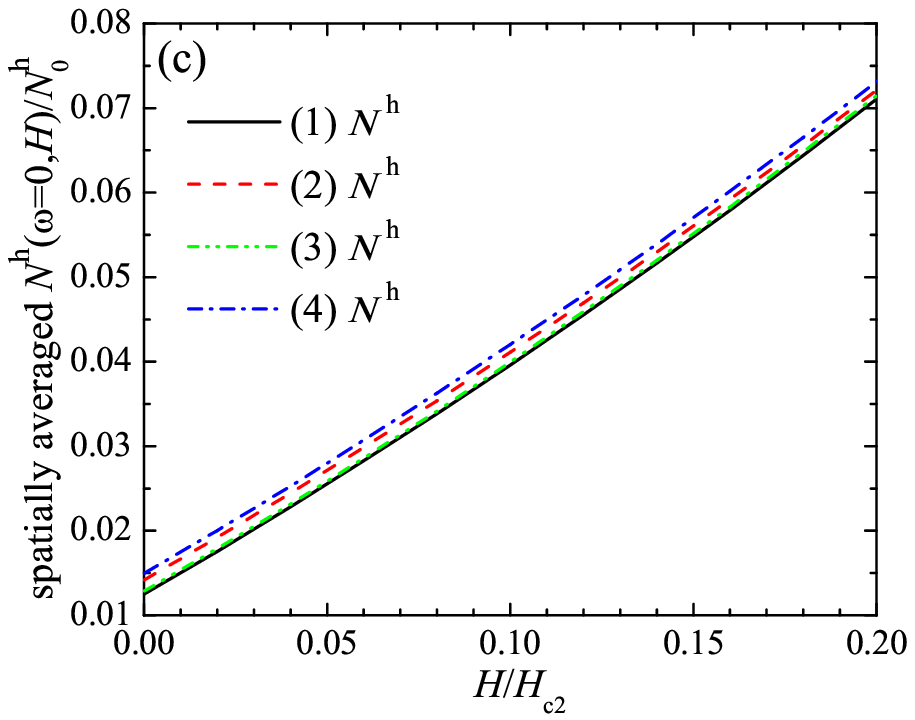}}
\caption{(Color online) Results of quasiclassical calculations for the parameters in Table I.(a) Magnetic
field dependence of the gaps in the two-band model calculated within the Pesch
approximation~\cite{Pesch_approx1,Pesch_approx2} for case (1) - (4). We assume $\Delta^e(H=0)=\Delta^h(H=0)$
here. The four sets of coupling constants $\lambda_{ij}$ are listed in TABLE \ref{tab:coupling_matrix}. (b)
Field dependence of the space average ZDOS $N^e(H)$ on the electron pocket for the four cases with
anisotropic gap with angular variation $\Phi_e(\theta)=(1+r\cos 2\theta)/\sqrt{1+r^2/2}$. (c) Field
dependence of the space average ZDOS $N^h(H)$ for the four cases with isotropic gap along the hole pocket.}
\label{fig:GapsH-SpaceAve}
\end{figure*}

To illustrate that the salient features of the vortex state DOS are captured in our approach in
Fig.~\ref{fig:onebandXi} we show the field dependence of the spatially averaged zero energy local density of
states (ZDOS) for a one-band SC with either an isotropic $s$-wave gap or a strongly anisotropic nodal gap
($r=1.3$). Note that, while the field dependences in both the nodal and fully gapped cases clearly fit the
anticipated power laws at low fields, $\sqrt{H}$ and $H$, respectively, there is a significant influence on
the magnitude of the DOS caused by the size of the core, with the smaller core size yielding smaller ZDOS.
In particular, in the absence of the Kramer-Pesch effect, for the nodal case the ZDOS would exceed the normal
state value at fields far below $H_{c2}$, which is unphysical.

Below we consider $r=0.9$ and $r=1.3$ to mimic a gap with deep minima and accidental nodes, respectively. To
show different types of behavior allowed within our microscopic model we chose four sets of coupling
constants, two for each value of $r$, as shown in Table~\ref{tab:coupling_matrix}. In cases (1) and (3), the
interband pairing $\lambda_{12}$ is strong and close to the intraband parameter $\lambda_{11}$, while in case
(2) and (4) $\lambda_{12}\ll\lambda_{11},\lambda_{22}$.
\begin{table}
\caption{The different models for the coupling matrix and the gap
         anisotropy on the electron pockets considered in this work. }
\label{tab:coupling_matrix}

\renewcommand{\arraystretch}{1.4}
\begin{tabular*}{\columnwidth}{@{\extracolsep{\fill}}cccccccc}
\hline \hline
 & $\lambda_{11}$& $\lambda_{12}$ & $\lambda_{21}$ & $\lambda_{22}$ &$r$ &$T_c/$K &$H_{c2}/$T\\\hline
case (1)  & 0.51 & 0.51 & 0.33 & 0.65 & 0.9 & 31 & 54 \\
case (2)  & 1.00 & 0.02 & 0.013 & 0.81 & 0.9 & 31 & 47 \\
case (3)  & 0.51 & 0.51 & 0.34 & 0.64 & 1.3  & 31 & 54  \\
case (4)  & 1.00 & 0.023 & 0.015 & 0.77 & 1.3 & 31 & 42 \\
\hline \hline
\end{tabular*}
\end{table}

In Fig.~\ref{fig:GapsH-SpaceAve}(a) we show the self-consistently determined magnitudes of the bulk gaps in
the vortex state $\Delta_{1,2}(H)$ as defined in Eq.~(\ref{eq:gapEqn}) and (\ref{eq:gapforms}). $H_{c2}\sim
40-50$~T. In the cases with only weak interband pairing (2) and (4), the gap on the electron Fermi surface
deviates considerably from the phenomenological form $\Delta(H)=\Delta_0\sqrt{1-H/H_{c2}}$.
Figs.~\ref{fig:GapsH-SpaceAve}(b) and (c) show the spatially averaged ZDOS corresponding to each band. For
$N^e(H)$ and for $r=1.3$ the $\sqrt{H}$ behavior of the Volovik effect is clearly visible at lower fields.
Comparing Fig.~\ref{fig:GapsH-SpaceAve}(b) to Fig.~\ref{fig:onebandXi} we find that within the two-band model
the density of states on the electron band $N^e(H)$ reaches a quasi-linear behavior already at smaller fields
than the corresponding density of states for the one-band case. In Fig.~\ref{fig:onebandXi} a linear behavior
is never observed, and might only be fit over some intermediate field range  for $H/H_{c2}>0.2$, while in the
multiband case $N^e(H)$ displays a clear linear behavior already for $H/H_{c2}>0.1$.

It is tempting to interpret the low-field crossover to a quasilinear field variation  as evidence for a small
energy scale $\Delta_{sm}\equiv \Delta_0^e(1-r)/\sqrt{1+r^2/2}$ on the electron band; this, however, seems
unlikely. Provided $\Delta_{sm}\ll\Delta_0^e$, the gap still increases linearly along the Fermi surface away
from the nodal points above this energy scale, simply with a different slope. Then within the usual Volovik
argumentation the contributions from extended states at these intermediate energies give rise to a $\sqrt{H}$
contribution even if $\Delta_{sm}\lesssim E_H \ll \Delta_{max}$, where $E_H \propto \sqrt{H}$ is the average
Doppler shift and $\Delta_{max}\equiv \Delta_0^e(1+r)/\sqrt{1+r^2/2}$ is the maximum gap.  There is therefore
no true linear-$H$ behavior arising from the electron band with gap nodes. Consequently, we interpret this
crossover as the consequence of the two-band behavior coupled with a gradually increasing contribution of
core states which is nearly linear in field. Fig.~\ref{fig:GapsH-SpaceAve}(c) clearly shows that the density
of states on the hole band $N^h(H)$, assumed here to be fully gapped, is always linear as a function of field
and the results for the two different coupling matrices considered here are very similar. However, as
mentioned before, the slope is smaller than the one predicted for an idealized $s$-wave SC.

\begin{figure}
\includegraphics[width=0.45\textwidth]{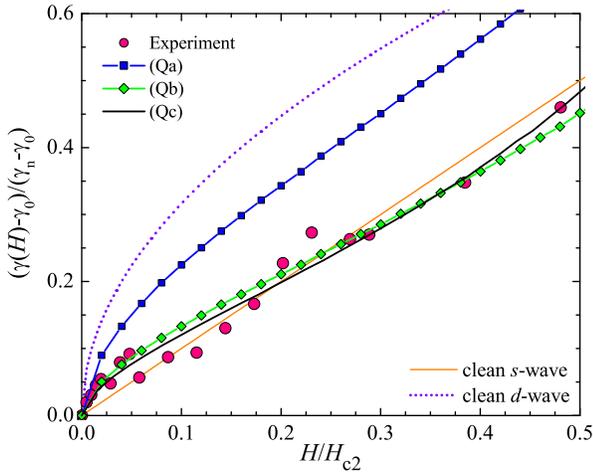}
\caption{\label{fig:totSpaceAve} (Color online) Comparison of the experimentally measured normalized specific
heat coefficient (large pink dots) to different theoretical results for the spatially averaged ZDOS. The
dotted violet and solid orange curves are the predictions for the spatially averaged ZDOS for a clean
$s$-wave and $d$-wave SC. The blue squares (case (Qa)) and green diamonds (case (Qb)) are the differently
weighted sums of $\bar{N}^e(H)$ and $\bar{N}^h(H)$ evaluated for case (4) of
Figs.~\ref{fig:GapsH-SpaceAve}(b) and (c). The black line (case (Qc)) is obtained using the formula
$\gamma^\text{tot}=a_1\bar{N}_e(H)+a_2\bar{N}_h(H)$ where $a_1=3.2$~mJ/(mole~K$^2$),
$a_2=10.3$~mJ/(mole~K$^2$) are determined with the least square fit to experimental data below 30~T. Note
``$d$-wave'' and ``$s$-wave'' curves represent simple extrapolations of the low-field $\sqrt{H}$ and $H$
terms up to $H_{c2}$.}
\end{figure}

Using Eq.~(\ref{eq:sumDOSeh}), the spatially averaged ZDOS on the electron and the hole band are added with
different weights. Using the results presented in Figs.~\ref{fig:GapsH-SpaceAve}(b) and (c) as case (4), we
investigate several cases. Since there are two electron pockets, and assuming that only one hole pocket
contributes significantly to the low energy density of states (or that a naive average over the hole pockets
is sufficient), the net DOS and the field dependence of the Sommerfeld coefficient are only functions of the
ratio of the densities of states on the electron and hole sheets.  In the following we will study three cases
that we will abbreviate with ``Q" indicating the use of the quasiclassical, or Eilenberger, approach:
\begin{itemize}
 \item (Qa): 
 we assume that only one hole pocket contributes
  considerably to the low energy DOS, and use the weights $w_e/w_h=2N_0^e/N_0^h$ taken from the DFT calculation,
  $N_0^e/N_0^h=0.65$, see  Ref.~\onlinecite{Mishra2011};
 \item (Qb): 
  We once again fix $N_0^e/N_0^h=0.65$, but adopt a model for which the normal DOS for all three hole pockets of
  Ba$_2$Fe$_2$(As$_{0.7}$P$_{0.3}$)$_2$ are the same and for which all three pockets contribute equally to the low energy
  DOS, hence $w_e/w_h=2N_0^e/3N_0^h$;
 \item (Qc): 
  We do not hold the ratio $N_0^e/N_0^h$ fixed, but instead calculate the weights for the electron pockets $a_1$
  and for the hole pockets $a_2$ by a least squares fit to the experimental data using the formula
  $\gamma^\text{tot}=a_1\bar{N}_e(H)+a_2\bar{N}_h(H)$. If we normalize it to the presumed contribution of the
  superconducting fraction,
  $\gamma_n-\gamma_0\approx 14$~mJ/(mole~K$^2$), where $\gamma_0$ is the extraneous term, see below, we find
  $w_e/(w_e+w_h)=a_1/(\gamma_n-\gamma_0)$
  and $w_h/(w_e+w_h)=a_2/(\gamma_n-\gamma_0)$ and $a_1/a_2=w_e/w_h$.
\end{itemize}
In Fig.~\ref{fig:totSpaceAve} we compare the results for all three cases to the experimentally measured
specific heat coefficient (pink dots). The experimental values are obtained by extrapolating the measured
specific heat coefficient $\gamma$ at various temperatures to $T=0$. The upper critical field $H_{c2}$ is
taken to be 52 T, see Ref.~\onlinecite{Yamashita11}. The normal state $\gamma_n=16$ mJ/(mol~K$^2$) can be
obtained by extrapolating $\gamma$ to $H_{c2}$. A substantial residual\cite{Kim10} $\gamma_0=1.7$
mJ/(mol~K$^2$) in the superconducting state, presumed due to disorder, is subtracted in the plots of the
field dependence from the experimental data (pink dots) to compare with our quasiclassical calculation in the
clean limit (blue squares and green diamonds). Note that subtracting of the residual C/T tends to enhance the
scatter in the low-T data of Fig 2.

From Fig.~\ref{fig:totSpaceAve}, we see that the results derived for model (Qb) with three equal mass hole
pockets and two equal mass electron pockets are in good agreement with the experimental data: both experiment
and theory show a ``Volovik effect" at the lowest fields and then a crossover to a linear $H$ dependence at
intermediate fields. While model (Qa) has the same qualitative behavior, the relative weights of hole and
electron bands are apparently not consistent with the normalized experimental data, and the fit is much
poorer. Compared to (Qb) the least squares fit (Qc) to the experimental data (black line) is only marginally
improved, and gives $N_0^e/N_0^h=0.47$ with two electron pockets/three hole pockets or $0.16$ with two
electron pockets/one hole pocket, same order as obtained from DFT.

\begin{figure}
\includegraphics[width=0.45\textwidth]{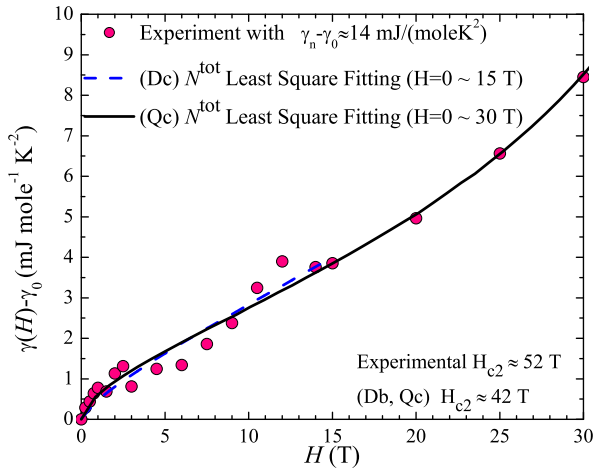}
\caption{\label{fig:BestFitQD} (Color online) $\gamma^\text{tot}$ plotted with experimental data in absolute
units. Case (Qc) is the similar black line in Fig.~\ref{fig:totSpaceAve}. (Dc) is obtained using formula
$\gamma^\text{tot}=a_1\bar{N}_e(H)+a_2\bar{N}_h(H)$, where $\bar{N}_e(H)$ and $\bar{N}_h(H)$ are represented
by the open squares and circles in Fig.~\ref{fig:DopplerSpaceAve} and $a_1=1.50$~mJ/(mole~K$^2$),
$a_2=65.6$~mJ/(mole~K$^2$) for (Dc) are determined with the least square fit to the experimental data below
15~T.}
\end{figure}

For completeness it is important to determine whether the experimental data can be appropriately fit within
the confines of a simple two-band Doppler shift approach. We detail this method in the Appendix, where we
show that models (a) and (b) do not give a satisfactory fit to the experiment. In contrast, model (c) yields
a rather similar field dependence of the field-induced enhancement of the Sommerfeld coefficient for the
quasiclassical and Doppler (Dc) methods, as shown in Fig.~\ref{fig:BestFitQD}. At the same time the best fit
linear coefficients for (Dc), $a_1=1.50$~mJ/(mole~K$^2$), $a_2=65.6$~mJ/(mole~K$^2$), give the ratio of the
normal state DOS for two electron/three hole Fermi sheets of $N_0^e/N_0^h=3a_1/2a_2\approx 0.03$, very
different from the value of 0.65 obtained within the band structure calculations. Consequently, the
quasiclassical methods provides a far more satisfying fit to the data.

\begin{figure}
\includegraphics[width=0.45\textwidth]{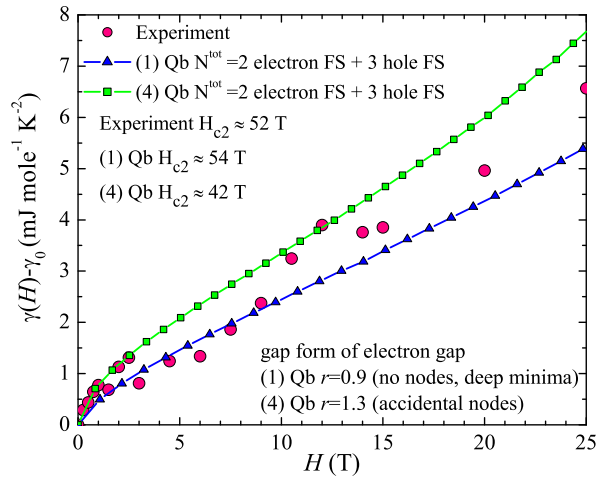}
\caption{\label{fig:exptcase1and4}  (Color online) Experimentally measured specific heat coefficient (large
pink dots) compared to calculations with deep gap minima (case (1), $r=0.9$, blue triangles) and accidental
nodes (case (4), $r=1.3$, green squares). In both cases the weight of electron and hole pocket contributions
has been chosen in agreement with case (Qb).}
\end{figure}

As is usually the case with the measurements that probe the amplitude rather than the phase of the gap, it is
difficult to distinguish the deep minima from the nodes. In this case we find that with our current
uncertainty in the band parameters, and the scatter in the data, it is impossible to assert the nodal
behavior purely from the current data.  Fig.~\ref{fig:exptcase1and4} shows the comparison of cases (1) and
case (4) of Table~\ref{tab:coupling_matrix}, corresponding to $r=1.3$ and 0.9, i.e. with and without true
nodes, with the weights of case (Qb). Even though the nodal fit appears better at the lowest fields, higher
$H$ data are in between the two cases. Therefore the conclusion about the true node comes from the data on
other experiments, such as penetration depth.

\section{Conclusions}\label{sec:conclusions}

Among the various families of Fe-based superconductors, \BFAP may be a key system for understanding the
origins of superconductivity. In part this is because, alone among the materials thought to display nodes in
the superconducting gap, it possesses a rather high $T_c$ of 31~K, and hence the interplay of the pairing
mechanism and Fermi surface shape and parameters in determining the gap anisotropy is under special scrutiny.
The lack of an observable Volovik effect in earlier specific heat measurements was a cautionary note in an
otherwise consistent array of measurements in support of gap nodes.  In this paper, we have presented new
experimental data at both lower and higher fields than previous measurements, and found that the initially
reported linear-$H$ behavior extends up to 35~T, but that at low fields ($H\lesssim 4$ T) more precise
measurements with smaller gradations in the change of field between data points are now clearly consistent
with a Volovik-type effect. The residual $T\to 0$ Sommerfeld coefficient $\gamma(T\to 0)$ is about 1.7
mJ/mol-K$^2$, consistent with possible nanoscale disorder in the sample.  The low-field sublinear dependence
of the Sommerfeld coefficient is a strong indication that nodes (or deep minima) are present, and provides
the sought-after consistency with other probes without having to make extreme assumptions about the ratio of
masses on electron pockets to those on hole pockets, as was proposed in Ref.~\onlinecite{Kim10}.

It is nevertheless striking that indications of nodal behavior on the same samples is so much weaker in the
specific heat measurements as compared to thermal conductivity and penetration depth. This is clearly
indicating that the nodes are located on the pockets with smaller masses and/or longer lifetimes, as was
pointed out in Ref.~\onlinecite{Kim10}. We have attempted to put this statement on a semiquantitative basis
by presenting a quasiclassical (Eilenberger) calculation of the density of states and specific heat of a
two-band anisotropic $s_\pm$ superconductor.  Comparison with the Doppler shift method allowed us to argue
that the quasiclassical calculation is superior for semiquantitative purposes.  We find that the unusually
small range of Volovik-type behavior, followed by a large range of linear-$H$ behavior, is due to the small
gap and weak nodes on the small mass (presumably electron) sheet.\cite{Kim10,Yamashita11} Good fits to the
data are obtained for average hole and electron maximum gaps of approximately equal magnitude, in the weak
interband coupling limit. The success of this fit should not, however, tempt one to draw definitive
conclusions about the relative magnitudes of the pairing interactions. The proliferation of parameters in the
theory make it difficult to determine gap magnitudes, density of states ratios, and nodal properties with any
quantitative certainty.  Equally good fits can be obtained, for example, with substantially smaller full gaps
than anisotropic gaps; the nodes control the low-field behavior, and the small full gap gives rise to a large
linear term.  What is important is that we have shown that a fit can be obtained, with reasonable values of
the parameters, that it can only be obtained if nodes exist on one of the Fermi sheets, and that it requires
going beyond the simple Doppler shift picture. It is our hope that the results of this calculation and fit
will eventually lead to a more quantitative first principles based calculation.

\begin{acknowledgments}
The authors are grateful to F. Ronning for useful discussions. YW and PJH were supported by the DOE under
DE-FG02-05ER46236, and GRS and JSK  under DE-FG02-86ER45268. I. V. acknowledges support from DOE Grant
DE-FG02-08ER46492. SG, PH, YM, TS, and IV are grateful to the Kavli Institute of Theoretical Physics for its
support and hospitality during the research and writing of this paper.
\end{acknowledgments}

\appendix

\section{Comparison with the Doppler-shift method}
In the following we briefly discuss the basic concepts of the Doppler-shift method and compare it to the
quasiclassical approximation as manifested in the formulation of the Eilenberger equations introduced in the
main text. The Doppler shifted energy due to the local supercurrent flow is
$\omega-m\mathbf{v}_F\cdot\mathbf{v}_s(\mathbf{r})$ where
\begin{align}
  m\mathbf{v}_F^{e,h}\cdot\mathbf{v}_s(\mathbf{r})
    &=\frac{\hbar}{2|\mathbf{r}|}\mathbf{v}_F^{e,h}(\theta)\cdot\mathbf{e}_\phi
    =\frac{\hbar v_F^{e,h}}{2|\mathbf{r}|}\sin(\theta-\phi) \notag\\
    &=\frac{\Delta_{0}^{e,h}}{2\tilde{\rho}^{e,h}}\sin(\theta-\phi).
\end{align}
Here, we assume $\Delta^e(T=0,H=0)=\Delta^h(T=0,H=0)=\Delta_0^{e,h}$ and use
$\tilde{\rho}^{e,h}=|\mathbf{r}|/\xi_0^{e,h}$. Therefore the normalized local DOS is
\begin{align}
  &N(\omega,\mathbf{r})=\int_0^{2\pi}\frac{d\theta}{2\pi}\notag\\
  &\mathrm{Re}\,\left\{
    \frac{|\omega-m\mathbf{v}_F\cdot\mathbf{v}_s(\mathbf{r})|}
    {\sqrt{(\omega-m\mathbf{v}_F\cdot\mathbf{v}_s(\mathbf{r}))^2-|\Delta(\theta)|^2}}
    \right\},
\end{align}
and thus the normalized spatially averaged DOS is
\begin{align}
  &\bar{N}(H)=\int_0^{\tilde{R}} \frac{d\tilde{\rho} \; \tilde{\rho}}{\pi \tilde{R}^2} \int_0^{2\pi} d\phi
    \int_0^{2\pi}\frac{d\theta}{2\pi} \notag\\
    &\mathrm{Re}
    \left\{\frac{|\omega-\Delta_{0}\sin(\theta-\phi)/(2\tilde{\rho})|}
    {\sqrt{\left[\omega-\Delta_{0}\sin(\theta-\phi)/(2\tilde{\rho})\right]^2
    -|\Delta(\theta)|^2}}\right\}.
\end{align}
Here we have introduced the normalized vortex cell radius $\tilde{R}=R/\xi_0^{e}$ for the electron bands or
$R/\xi_0^{h}$ for the hole bands, respectively. Since the Doppler-shift method does not capture core state
contributions it underestimates the slope of the magnetic field dependence of the zero energy DOS of an
$s$-wave SC. Since the core region only gives negligible contributions to the total DOS one can in principle
avoid the divergence of the Doppler-shift energy as $\tilde{\rho} \rightarrow 0$ by cutting out the complete
core region with a lower limit $\xi_0$ for the radial integration. Here we have included the core region when
integrating over the vortex unit cell. To model $\Delta(\theta)$ we use a similar function as given by
Eq.~(\ref{eq:gapforms}), but without explicitely modeling the core structure
\begin{subequations}
\begin{align}
  \Delta^{e}(\vec{\rho};\theta)&=\Delta_1(H=0)
      \frac{1+r\cos2\theta}{\sqrt{1+r^2/2}},\\
  \Delta^{h}(\vec{\rho})&=\Delta_2(H=0),
\end{align}
\end{subequations}
and we use the self-consistently calculated $\Delta_{1,2}(H=0)$ in case (4) of the quasiclassical calculation
in which the anisotropy factor $r=1.3$ for the gap along the electron Fermi surface sheet and the ratio of
the normal DOS at the Fermi energy is taken as $N_0^e/N_0^h=0.65$. 
\begin{figure}
\includegraphics[width=0.48\textwidth]{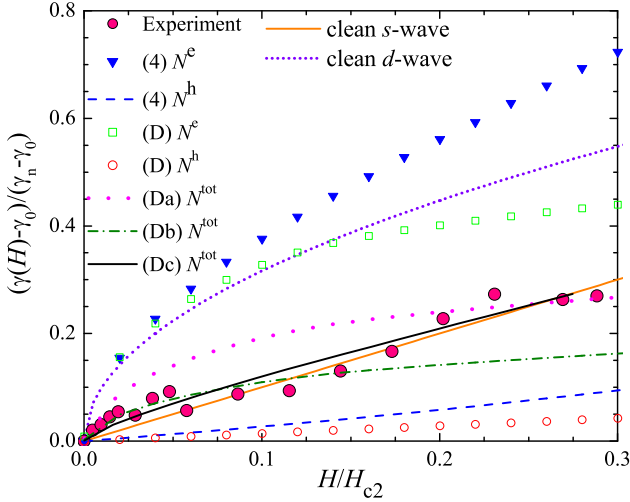}
\caption{\label{fig:DopplerSpaceAve} (Color online) Normalized specific heat coefficient determined within
the quasiclassical approach (blue triangles and blue dashed curve) and within the Doppler-shift method (open
squares and circles). Total normalized specific heat coefficients obtained within Doppler-shift method are
shown by magenta dots, olive dash-dots and black solid line for cases (Da, Db, Dc), corresponding to (Qa, Qb,
Qc), respectively. The dotted violet and solid orange curves are the power laws predicted for the spatially
averaged ZDOS of an idealized $s$-wave and $d$-wave SC. Large pink dots show the experimentally determined
normalized specific heat coefficient.}
\end{figure}
In Fig.~\ref{fig:DopplerSpaceAve} we show the results obtained within the Doppler-shift approach and compare
them to case (4) of the quasiclassical method. Again we also show predictions for an idealized clean $s$- and
$d$-wave SC. We conclude that the Doppler-shift method and the quasiclassical method give comparable results
at lowest fields but start to deviate as soon as the field increases. One reason might be that with
increasing field and decreasing inter-vortex distance the core states that are not correctly accounted for
within the Doppler-shift method but captured within the quasiclassical approach become increasingly more
important. However, due to the limitations of the single vortex approximation the overlapping of vortices is
not correctly reproduced and the DOS is overestimated (In Fig.~\ref{fig:DopplerSpaceAve} the blue triangles
rise too fast).

In Fig.~\ref{fig:BestFitQD} we compare the least squares fit by Doppler-shift method (blue dashed curve)
together with least square fit by quasiclassical method (the similar black line from
Fig.~\ref{fig:totSpaceAve}) and experimental data (large pink dots). The linear coefficients for (Dc) are
$a_1=1.50$~mJ/(mole~K$^2$), $a_2=65.6$~mJ/(mole~K$^2$). Compared to the linear coefficients for (Qc)
($a_1=3.2$~mJ/(mole~K$^2$), $a_2=10.3$~mJ/(mole~K$^2$)), they give a nonphysical ratio of the normal DOS at
Fermi energy if we consider two electron Fermi sheets and three hole Fermi sheets. To see this point, let's
consider equation
\begin{align}
  C_{e,h}(T,H)&=\frac{1}{2}\int_{-\infty}^{+\infty}d\omega
  \frac{\omega^2\tilde{N}_{e,h}(H,\omega)}{T^2\cosh(\omega/2T)}\\
  &\approx A\tilde{N}_{e,h}(H,\omega)T  \quad\quad (\text{as $T\to 0$}) \notag\\
  \Rightarrow \gamma_{e,h}&\approx A\tilde{N}_{e,h}(H,0)\equiv AN_0^{e,h}\bar{N}_{e,h}(H,0).\notag
\end{align}
Here $A$ is a numeric constant and we write $\tilde{N}_{e,h}(H,0)\equiv N_0^{e,h}\bar{N}_{e,h}(H,0)$ where
$\bar{N}_{e,h}(H,0)$ is the ZDOS calculated by the Green's function method as defined in
Eq.~(\ref{eq:dosHInt}). Denote the number of Fermi sheets included in summation as $n_{FS}$ and define
$n_{FS}^e A N_0^e=a_1$ and $n_{FS}^h A N_0^h=a_2$ (equivalent to $w_e=n_{FS}^e N_0^e$ and $w_h=n_{FS}^h
N_0^h$). Therefore $\gamma^\text{tot}=a_1\bar{N}_e(H)+a_2\bar{N}_h(H)$. Note $\bar{N}_{e,h}(H)$ are
dimensionless and $a_{1,2}$ are in unit of $\mathrm{mJ/(mole\,K^2)}$. Optimized parameters $a_{1,2}$ for
least square fit (Dc) to the experimental data below 15~T are $a_1=1.50$ and $a_2=65.6$. This leads to our
estimate in the main text of $a_1/a_2=(n_{FS}^e A N_0^e)/(n_{FS}^h A N_0^h)=(2N_0^e)/(3N_0^h)$, and
$N_0^e/N_0^h=0.03$, and to our conclusion that Doppler-shift method does not provide a satisfying physical
explanation to our specific heat experiment.


\begin{thebibliography}{44}%
\makeatletter
\providecommand \@ifxundefined [1]{%
 \@ifx{#1\undefined}
}%
\providecommand \@ifnum [1]{%
 \ifnum #1\expandafter \@firstoftwo
 \else \expandafter \@secondoftwo
 \fi
}%
\providecommand \@ifx [1]{%
 \ifx #1\expandafter \@firstoftwo
 \else \expandafter \@secondoftwo
 \fi
}%
\providecommand \natexlab [1]{#1}%
\providecommand \enquote  [1]{``#1''}%
\providecommand \bibnamefont  [1]{#1}%
\providecommand \bibfnamefont [1]{#1}%
\providecommand \citenamefont [1]{#1}%
\providecommand \href@noop [0]{\@secondoftwo}%
\providecommand \href [0]{\begingroup \@sanitize@url \@href}%
\providecommand \@href[1]{\@@startlink{#1}\@@href}%
\providecommand \@@href[1]{\endgroup#1\@@endlink}%
\providecommand \@sanitize@url [0]{\catcode `\\12\catcode `\$12\catcode
  `\&12\catcode `\#12\catcode `\^12\catcode `\_12\catcode `\%12\relax}%
\providecommand \@@startlink[1]{}%
\providecommand \@@endlink[0]{}%
\providecommand \url  [0]{\begingroup\@sanitize@url \@url }%
\providecommand \@url [1]{\endgroup\@href {#1}{\urlprefix }}%
\providecommand \urlprefix  [0]{URL }%
\providecommand \Eprint [0]{\href }%
\providecommand \doibase [0]{http://dx.doi.org/}%
\providecommand \selectlanguage [0]{\@gobble}%
\providecommand \bibinfo  [0]{\@secondoftwo}%
\providecommand \bibfield  [0]{\@secondoftwo}%
\providecommand \translation [1]{[#1]}%
\providecommand \BibitemOpen [0]{}%
\providecommand \bibitemStop [0]{}%
\providecommand \bibitemNoStop [0]{.\EOS\space}%
\providecommand \EOS [0]{\spacefactor3000\relax}%
\providecommand \BibitemShut  [1]{\csname bibitem#1\endcsname}%
\let\auto@bib@innerbib\@empty
\bibitem [{\citenamefont {Kamihara}\ \emph {et~al.}(2008)\citenamefont
  {Kamihara}, \citenamefont {Watanabe}, \citenamefont {Hirano},\ and\
  \citenamefont {Hosono}}]{kamihara08}%
  \BibitemOpen
  \bibfield  {author} {\bibinfo {author} {\bibfnamefont {Y.}~\bibnamefont
  {Kamihara}}, \bibinfo {author} {\bibfnamefont {T.}~\bibnamefont {Watanabe}},
  \bibinfo {author} {\bibfnamefont {M.}~\bibnamefont {Hirano}}, \ and\ \bibinfo
  {author} {\bibfnamefont {H.}~\bibnamefont {Hosono}},\ }\href@noop {}
  {\bibfield  {journal} {\bibinfo  {journal} {Journal of the American Chemical
  Society}\ }\textbf {\bibinfo {volume} {130}},\ \bibinfo {pages} {3296}
  (\bibinfo {year} {2008})}\BibitemShut {NoStop}%
\bibitem [{\citenamefont {Hsu}\ \emph {et~al.}(2008)\citenamefont {Hsu},
  \citenamefont {Luo}, \citenamefont {Yeh}, \citenamefont {Chen}, \citenamefont
  {Huang}, \citenamefont {Wu}, \citenamefont {Lee}, \citenamefont {Huang},
  \citenamefont {Chu}, \citenamefont {Yan} \emph {et~al.}}]{Hsu08}%
  \BibitemOpen
  \bibfield  {author} {\bibinfo {author} {\bibfnamefont {F.}~\bibnamefont
  {Hsu}}, \bibinfo {author} {\bibfnamefont {J.}~\bibnamefont {Luo}}, \bibinfo
  {author} {\bibfnamefont {K.}~\bibnamefont {Yeh}}, \bibinfo {author}
  {\bibfnamefont {T.}~\bibnamefont {Chen}}, \bibinfo {author} {\bibfnamefont
  {T.}~\bibnamefont {Huang}}, \bibinfo {author} {\bibfnamefont
  {P.}~\bibnamefont {Wu}}, \bibinfo {author} {\bibfnamefont {Y.}~\bibnamefont
  {Lee}}, \bibinfo {author} {\bibfnamefont {Y.}~\bibnamefont {Huang}}, \bibinfo
  {author} {\bibfnamefont {Y.}~\bibnamefont {Chu}}, \bibinfo {author}
  {\bibfnamefont {D.}~\bibnamefont {Yan}},  \emph {et~al.},\ }\href@noop {}
  {\bibfield  {journal} {\bibinfo  {journal} {Proceedings of the National
  Academy of Sciences}\ }\textbf {\bibinfo {volume} {105}},\ \bibinfo {pages}
  {14262} (\bibinfo {year} {2008})}\BibitemShut {NoStop}%
\bibitem [{\citenamefont {Wen}(2011)}]{Wen11}%
  \BibitemOpen
  \bibfield  {author} {\bibinfo {author} {\bibfnamefont {H.}~\bibnamefont
  {Wen}},\ }\href@noop {} {\bibfield  {journal} {\bibinfo  {journal} {Annual
  Review of Condensed Matter Physics}\ }\textbf {\bibinfo {volume} {2}},\
  \bibinfo {pages} {121} (\bibinfo {year} {2011})}\BibitemShut {NoStop}%
\bibitem [{\citenamefont {Stewart}(2011)}]{stewart11}%
  \BibitemOpen
  \bibfield  {author} {\bibinfo {author} {\bibfnamefont {G.}~\bibnamefont
  {Stewart}},\ }\href@noop {} {\bibfield  {journal} {\bibinfo  {journal}
  {arXiv:1106.1618}\ } (\bibinfo {year} {2011})}\BibitemShut {NoStop}%
\bibitem [{\citenamefont {Kemper}\ \emph {et~al.}(2010)\citenamefont {Kemper},
  \citenamefont {Maier}, \citenamefont {Graser}, \citenamefont {Cheng},
  \citenamefont {Hirschfeld},\ and\ \citenamefont {Scalapino}}]{Kemper10}%
  \BibitemOpen
  \bibfield  {author} {\bibinfo {author} {\bibfnamefont {A.}~\bibnamefont
  {Kemper}}, \bibinfo {author} {\bibfnamefont {T.}~\bibnamefont {Maier}},
  \bibinfo {author} {\bibfnamefont {S.}~\bibnamefont {Graser}}, \bibinfo
  {author} {\bibfnamefont {H.}~\bibnamefont {Cheng}}, \bibinfo {author}
  {\bibfnamefont {P.}~\bibnamefont {Hirschfeld}}, \ and\ \bibinfo {author}
  {\bibfnamefont {D.}~\bibnamefont {Scalapino}},\ }\href@noop {} {\bibfield
  {journal} {\bibinfo  {journal} {New Journal of Physics}\ }\textbf {\bibinfo
  {volume} {12}},\ \bibinfo {pages} {073030} (\bibinfo {year}
  {2010})}\BibitemShut {NoStop}%
\bibitem [{\citenamefont {Hirschfeld}\ \emph {et~al.}(2011)\citenamefont
  {Hirschfeld}, \citenamefont {Korshunov},\ and\ \citenamefont
  {Mazin}}]{HKMreview}%
  \BibitemOpen
  \bibfield  {author} {\bibinfo {author} {\bibfnamefont {P.}~\bibnamefont
  {Hirschfeld}}, \bibinfo {author} {\bibfnamefont {M.}~\bibnamefont
  {Korshunov}}, \ and\ \bibinfo {author} {\bibfnamefont {I.}~\bibnamefont
  {Mazin}},\ }\href@noop {} {\bibfield  {journal} {\bibinfo  {journal}
  {arXiv:1106.3712}\ } (\bibinfo {year} {2011})}\BibitemShut {NoStop}%
\bibitem [{\citenamefont {Kasahara}\ \emph {et~al.}(2010)\citenamefont
  {Kasahara}, \citenamefont {Shibauchi}, \citenamefont {Hashimoto},
  \citenamefont {Ikada}, \citenamefont {Tonegawa}, \citenamefont {Okazaki},
  \citenamefont {Shishido}, \citenamefont {Ikeda}, \citenamefont {Takeya},
  \citenamefont {Hirata}, \citenamefont {Terashima},\ and\ \citenamefont
  {Matsuda}}]{Kasahara10}%
  \BibitemOpen
  \bibfield  {author} {\bibinfo {author} {\bibfnamefont {S.}~\bibnamefont
  {Kasahara}}, \bibinfo {author} {\bibfnamefont {T.}~\bibnamefont {Shibauchi}},
  \bibinfo {author} {\bibfnamefont {K.}~\bibnamefont {Hashimoto}}, \bibinfo
  {author} {\bibfnamefont {K.}~\bibnamefont {Ikada}}, \bibinfo {author}
  {\bibfnamefont {S.}~\bibnamefont {Tonegawa}}, \bibinfo {author}
  {\bibfnamefont {R.}~\bibnamefont {Okazaki}}, \bibinfo {author} {\bibfnamefont
  {H.}~\bibnamefont {Shishido}}, \bibinfo {author} {\bibfnamefont
  {H.}~\bibnamefont {Ikeda}}, \bibinfo {author} {\bibfnamefont
  {H.}~\bibnamefont {Takeya}}, \bibinfo {author} {\bibfnamefont
  {K.}~\bibnamefont {Hirata}}, \bibinfo {author} {\bibfnamefont
  {T.}~\bibnamefont {Terashima}}, \ and\ \bibinfo {author} {\bibfnamefont
  {Y.}~\bibnamefont {Matsuda}},\ }\href {\doibase 10.1103/PhysRevB.81.184519}
  {\bibfield  {journal} {\bibinfo  {journal} {Phys. Rev. B}\ }\textbf {\bibinfo
  {volume} {81}},\ \bibinfo {pages} {184519} (\bibinfo {year}
  {2010})}\BibitemShut {NoStop}%
\bibitem [{\citenamefont {Jiang}\ \emph {et~al.}(2009)\citenamefont {Jiang},
  \citenamefont {Xing}, \citenamefont {Xuan}, \citenamefont {Wang},
  \citenamefont {Ren}, \citenamefont {Feng}, \citenamefont {Dai}, \citenamefont
  {Xu},\ and\ \citenamefont {Cao}}]{jiang09}%
  \BibitemOpen
  \bibfield  {author} {\bibinfo {author} {\bibfnamefont {S.}~\bibnamefont
  {Jiang}}, \bibinfo {author} {\bibfnamefont {H.}~\bibnamefont {Xing}},
  \bibinfo {author} {\bibfnamefont {G.}~\bibnamefont {Xuan}}, \bibinfo {author}
  {\bibfnamefont {C.}~\bibnamefont {Wang}}, \bibinfo {author} {\bibfnamefont
  {Z.}~\bibnamefont {Ren}}, \bibinfo {author} {\bibfnamefont {C.}~\bibnamefont
  {Feng}}, \bibinfo {author} {\bibfnamefont {J.}~\bibnamefont {Dai}}, \bibinfo
  {author} {\bibfnamefont {Z.}~\bibnamefont {Xu}}, \ and\ \bibinfo {author}
  {\bibfnamefont {G.}~\bibnamefont {Cao}},\ }\href@noop {} {\bibfield
  {journal} {\bibinfo  {journal} {Journal of Physics: Condensed Matter}\
  }\textbf {\bibinfo {volume} {21}},\ \bibinfo {pages} {382203} (\bibinfo
  {year} {2009})}\BibitemShut {NoStop}%
\bibitem [{\citenamefont {Shishido}\ \emph {et~al.}(2010)\citenamefont
  {Shishido}, \citenamefont {Bangura}, \citenamefont {Coldea}, \citenamefont
  {Tonegawa}, \citenamefont {Hashimoto}, \citenamefont {Kasahara},
  \citenamefont {Rourke}, \citenamefont {Ikeda}, \citenamefont {Terashima},
  \citenamefont {Settai}, \citenamefont {\ifmmode~\bar{O}\else \={O}\fi{}nuki},
  \citenamefont {Vignolles}, \citenamefont {Proust}, \citenamefont {Vignolle},
  \citenamefont {McCollam}, \citenamefont {Matsuda}, \citenamefont
  {Shibauchi},\ and\ \citenamefont {Carrington}}]{Shishido10}%
  \BibitemOpen
  \bibfield  {author} {\bibinfo {author} {\bibfnamefont {H.}~\bibnamefont
  {Shishido}}, \bibinfo {author} {\bibfnamefont {A.~F.}\ \bibnamefont
  {Bangura}}, \bibinfo {author} {\bibfnamefont {A.~I.}\ \bibnamefont {Coldea}},
  \bibinfo {author} {\bibfnamefont {S.}~\bibnamefont {Tonegawa}}, \bibinfo
  {author} {\bibfnamefont {K.}~\bibnamefont {Hashimoto}}, \bibinfo {author}
  {\bibfnamefont {S.}~\bibnamefont {Kasahara}}, \bibinfo {author}
  {\bibfnamefont {P.~M.~C.}\ \bibnamefont {Rourke}}, \bibinfo {author}
  {\bibfnamefont {H.}~\bibnamefont {Ikeda}}, \bibinfo {author} {\bibfnamefont
  {T.}~\bibnamefont {Terashima}}, \bibinfo {author} {\bibfnamefont
  {R.}~\bibnamefont {Settai}}, \bibinfo {author} {\bibfnamefont
  {Y.}~\bibnamefont {\ifmmode~\bar{O}\else \={O}\fi{}nuki}}, \bibinfo {author}
  {\bibfnamefont {D.}~\bibnamefont {Vignolles}}, \bibinfo {author}
  {\bibfnamefont {C.}~\bibnamefont {Proust}}, \bibinfo {author} {\bibfnamefont
  {B.}~\bibnamefont {Vignolle}}, \bibinfo {author} {\bibfnamefont
  {A.}~\bibnamefont {McCollam}}, \bibinfo {author} {\bibfnamefont
  {Y.}~\bibnamefont {Matsuda}}, \bibinfo {author} {\bibfnamefont
  {T.}~\bibnamefont {Shibauchi}}, \ and\ \bibinfo {author} {\bibfnamefont
  {A.}~\bibnamefont {Carrington}},\ }\href {\doibase
  10.1103/PhysRevLett.104.057008} {\bibfield  {journal} {\bibinfo  {journal}
  {Phys. Rev. Lett.}\ }\textbf {\bibinfo {volume} {104}},\ \bibinfo {pages}
  {057008} (\bibinfo {year} {2010})}\BibitemShut {NoStop}%
\bibitem [{\citenamefont {Hashimoto}\ \emph {et~al.}(2010)\citenamefont
  {Hashimoto}, \citenamefont {Yamashita}, \citenamefont {Kasahara},
  \citenamefont {Senshu}, \citenamefont {Nakata}, \citenamefont {Tonegawa},
  \citenamefont {Ikada}, \citenamefont {Serafin}, \citenamefont {Carrington},
  \citenamefont {Terashima}, \citenamefont {Ikeda}, \citenamefont {Shibauchi},\
  and\ \citenamefont {Matsuda}}]{Hashimoto10}%
  \BibitemOpen
  \bibfield  {author} {\bibinfo {author} {\bibfnamefont {K.}~\bibnamefont
  {Hashimoto}}, \bibinfo {author} {\bibfnamefont {M.}~\bibnamefont
  {Yamashita}}, \bibinfo {author} {\bibfnamefont {S.}~\bibnamefont {Kasahara}},
  \bibinfo {author} {\bibfnamefont {Y.}~\bibnamefont {Senshu}}, \bibinfo
  {author} {\bibfnamefont {N.}~\bibnamefont {Nakata}}, \bibinfo {author}
  {\bibfnamefont {S.}~\bibnamefont {Tonegawa}}, \bibinfo {author}
  {\bibfnamefont {K.}~\bibnamefont {Ikada}}, \bibinfo {author} {\bibfnamefont
  {A.}~\bibnamefont {Serafin}}, \bibinfo {author} {\bibfnamefont
  {A.}~\bibnamefont {Carrington}}, \bibinfo {author} {\bibfnamefont
  {T.}~\bibnamefont {Terashima}}, \bibinfo {author} {\bibfnamefont
  {H.}~\bibnamefont {Ikeda}}, \bibinfo {author} {\bibfnamefont
  {T.}~\bibnamefont {Shibauchi}}, \ and\ \bibinfo {author} {\bibfnamefont
  {Y.}~\bibnamefont {Matsuda}},\ }\href {\doibase 10.1103/PhysRevB.81.220501}
  {\bibfield  {journal} {\bibinfo  {journal} {Phys. Rev. B}\ }\textbf {\bibinfo
  {volume} {81}},\ \bibinfo {pages} {220501} (\bibinfo {year}
  {2010})}\BibitemShut {NoStop}%
\bibitem [{\citenamefont {Nakai}\ \emph {et~al.}(2010)\citenamefont {Nakai},
  \citenamefont {Iye}, \citenamefont {Kitagawa}, \citenamefont {Ishida},
  \citenamefont {Kasahara}, \citenamefont {Shibauchi}, \citenamefont
  {Matsuda},\ and\ \citenamefont {Terashima}}]{Nakai10}%
  \BibitemOpen
  \bibfield  {author} {\bibinfo {author} {\bibfnamefont {Y.}~\bibnamefont
  {Nakai}}, \bibinfo {author} {\bibfnamefont {T.}~\bibnamefont {Iye}}, \bibinfo
  {author} {\bibfnamefont {S.}~\bibnamefont {Kitagawa}}, \bibinfo {author}
  {\bibfnamefont {K.}~\bibnamefont {Ishida}}, \bibinfo {author} {\bibfnamefont
  {S.}~\bibnamefont {Kasahara}}, \bibinfo {author} {\bibfnamefont
  {T.}~\bibnamefont {Shibauchi}}, \bibinfo {author} {\bibfnamefont
  {Y.}~\bibnamefont {Matsuda}}, \ and\ \bibinfo {author} {\bibfnamefont
  {T.}~\bibnamefont {Terashima}},\ }\href {\doibase 10.1103/PhysRevB.81.020503}
  {\bibfield  {journal} {\bibinfo  {journal} {Phys. Rev. B}\ }\textbf {\bibinfo
  {volume} {81}},\ \bibinfo {pages} {020503} (\bibinfo {year}
  {2010})}\BibitemShut {NoStop}%
\bibitem [{\citenamefont {Yamashita}\ \emph {et~al.}(2011)\citenamefont
  {Yamashita}, \citenamefont {Senshu}, \citenamefont {Shibauchi}, \citenamefont
  {Kasahara}, \citenamefont {Hashimoto}, \citenamefont {Watanabe},
  \citenamefont {Ikeda}, \citenamefont {Terashima}, \citenamefont {Vekhter},
  \citenamefont {Vorontsov},\ and\ \citenamefont {Matsuda}}]{Yamashita11}%
  \BibitemOpen
  \bibfield  {author} {\bibinfo {author} {\bibfnamefont {M.}~\bibnamefont
  {Yamashita}}, \bibinfo {author} {\bibfnamefont {Y.}~\bibnamefont {Senshu}},
  \bibinfo {author} {\bibfnamefont {T.}~\bibnamefont {Shibauchi}}, \bibinfo
  {author} {\bibfnamefont {S.}~\bibnamefont {Kasahara}}, \bibinfo {author}
  {\bibfnamefont {K.}~\bibnamefont {Hashimoto}}, \bibinfo {author}
  {\bibfnamefont {D.}~\bibnamefont {Watanabe}}, \bibinfo {author}
  {\bibfnamefont {H.}~\bibnamefont {Ikeda}}, \bibinfo {author} {\bibfnamefont
  {T.}~\bibnamefont {Terashima}}, \bibinfo {author} {\bibfnamefont
  {I.}~\bibnamefont {Vekhter}}, \bibinfo {author} {\bibfnamefont {A.~B.}\
  \bibnamefont {Vorontsov}}, \ and\ \bibinfo {author} {\bibfnamefont
  {Y.}~\bibnamefont {Matsuda}},\ }\href {\doibase 10.1103/PhysRevB.84.060507}
  {\bibfield  {journal} {\bibinfo  {journal} {Phys. Rev. B}\ }\textbf {\bibinfo
  {volume} {84}},\ \bibinfo {pages} {060507} (\bibinfo {year}
  {2011})}\BibitemShut {NoStop}%
\bibitem [{\citenamefont {Kim}\ \emph {et~al.}(2010)\citenamefont {Kim},
  \citenamefont {Hirschfeld}, \citenamefont {Stewart}, \citenamefont
  {Kasahara}, \citenamefont {Shibauchi}, \citenamefont {Terashima},\ and\
  \citenamefont {Matsuda}}]{Kim10}%
  \BibitemOpen
  \bibfield  {author} {\bibinfo {author} {\bibfnamefont {J.~S.}\ \bibnamefont
  {Kim}}, \bibinfo {author} {\bibfnamefont {P.~J.}\ \bibnamefont {Hirschfeld}},
  \bibinfo {author} {\bibfnamefont {G.~R.}\ \bibnamefont {Stewart}}, \bibinfo
  {author} {\bibfnamefont {S.}~\bibnamefont {Kasahara}}, \bibinfo {author}
  {\bibfnamefont {T.}~\bibnamefont {Shibauchi}}, \bibinfo {author}
  {\bibfnamefont {T.}~\bibnamefont {Terashima}}, \ and\ \bibinfo {author}
  {\bibfnamefont {Y.}~\bibnamefont {Matsuda}},\ }\href {\doibase
  10.1103/PhysRevB.81.214507} {\bibfield  {journal} {\bibinfo  {journal} {Phys.
  Rev. B}\ }\textbf {\bibinfo {volume} {81}},\ \bibinfo {pages} {214507}
  (\bibinfo {year} {2010})}\BibitemShut {NoStop}%
\bibitem [{Note1()}]{Note1}%
  \BibitemOpen
  \bibinfo {note} {In contrast to BaFe$_2$(As$_{1-x}$P$_x$)$_2$, recent high
  field measurements on underdoped ($x=0.045$) and overdoped ($x=0.103$)
  BaFe$_{2-x}$Co$_x$As$_2$ have found that the specific heat coefficient varies
  approximately as $H^{0.7}$ all the way up to $H_{c2}(0)$. J. S. Kim, G. R.
  Stewart, K. Gofryk, F. Ronning, and A. S. Sefat, to be
  published.}\BibitemShut {Stop}%
\bibitem [{\citenamefont {Bang}(2010)}]{Bang10}%
  \BibitemOpen
  \bibfield  {author} {\bibinfo {author} {\bibfnamefont {Y.}~\bibnamefont
  {Bang}},\ }\href {\doibase 10.1103/PhysRevLett.104.217001} {\bibfield
  {journal} {\bibinfo  {journal} {Phys. Rev. Lett.}\ }\textbf {\bibinfo
  {volume} {104}},\ \bibinfo {pages} {217001} (\bibinfo {year}
  {2010})}\BibitemShut {NoStop}%
\bibitem [{\citenamefont {Kogan}\ and\ \citenamefont
  {Schmalian}(2011)}]{Kogan:2011}%
  \BibitemOpen
  \bibfield  {author} {\bibinfo {author} {\bibfnamefont {V.~G.}\ \bibnamefont
  {Kogan}}\ and\ \bibinfo {author} {\bibfnamefont {J.}~\bibnamefont
  {Schmalian}},\ }\href {\doibase 10.1103/PhysRevB.83.054515} {\bibfield
  {journal} {\bibinfo  {journal} {Phys. Rev. B}\ }\textbf {\bibinfo {volume}
  {83}},\ \bibinfo {pages} {054515} (\bibinfo {year} {2011})}\BibitemShut
  {NoStop}%
\bibitem [{\citenamefont {Stewart}(1983)}]{stewart83}%
  \BibitemOpen
  \bibfield  {author} {\bibinfo {author} {\bibfnamefont {G.}~\bibnamefont
  {Stewart}},\ }\href@noop {} {\bibfield  {journal} {\bibinfo  {journal}
  {Review of Scientific Instruments}\ }\textbf {\bibinfo {volume} {54}},\
  \bibinfo {pages} {1} (\bibinfo {year} {1983})}\BibitemShut {NoStop}%
\bibitem [{\citenamefont {Andraka}\ \emph {et~al.}(1989)\citenamefont
  {Andraka}, \citenamefont {Fraunberger}, \citenamefont {Kim}, \citenamefont
  {Quitmann},\ and\ \citenamefont {Stewart}}]{Andraka89}%
  \BibitemOpen
  \bibfield  {author} {\bibinfo {author} {\bibfnamefont {B.}~\bibnamefont
  {Andraka}}, \bibinfo {author} {\bibfnamefont {G.}~\bibnamefont
  {Fraunberger}}, \bibinfo {author} {\bibfnamefont {J.~S.}\ \bibnamefont
  {Kim}}, \bibinfo {author} {\bibfnamefont {C.}~\bibnamefont {Quitmann}}, \
  and\ \bibinfo {author} {\bibfnamefont {G.~R.}\ \bibnamefont {Stewart}},\
  }\href {\doibase 10.1103/PhysRevB.39.6420} {\bibfield  {journal} {\bibinfo
  {journal} {Phys. Rev. B}\ }\textbf {\bibinfo {volume} {39}},\ \bibinfo
  {pages} {6420} (\bibinfo {year} {1989})}\BibitemShut {NoStop}%
\bibitem [{\citenamefont {Kim}\ \emph {et~al.}(2009)\citenamefont {Kim},
  \citenamefont {Kim},\ and\ \citenamefont {Stewart}}]{Kim09}%
  \BibitemOpen
  \bibfield  {author} {\bibinfo {author} {\bibfnamefont {J.}~\bibnamefont
  {Kim}}, \bibinfo {author} {\bibfnamefont {E.}~\bibnamefont {Kim}}, \ and\
  \bibinfo {author} {\bibfnamefont {G.}~\bibnamefont {Stewart}},\ }\href@noop
  {} {\bibfield  {journal} {\bibinfo  {journal} {Journal of Physics: Condensed
  Matter}\ }\textbf {\bibinfo {volume} {21}},\ \bibinfo {pages} {252201}
  (\bibinfo {year} {2009})}\BibitemShut {NoStop}%
\bibitem [{\citenamefont {Eilenberger}(1968)}]{eilenb68}%
  \BibitemOpen
  \bibfield  {author} {\bibinfo {author} {\bibfnamefont {G.}~\bibnamefont
  {Eilenberger}},\ }\href@noop {} {\bibfield  {journal} {\bibinfo  {journal}
  {Z. Phys.}\ }\textbf {\bibinfo {volume} {214}},\ \bibinfo {pages} {195}
  (\bibinfo {year} {1968})}\BibitemShut {NoStop}%
\bibitem [{\citenamefont {Larkin}\ and\ \citenamefont
  {Ovchinnikov}(1969)}]{larkin69}%
  \BibitemOpen
  \bibfield  {author} {\bibinfo {author} {\bibfnamefont {A.~I.}\ \bibnamefont
  {Larkin}}\ and\ \bibinfo {author} {\bibfnamefont {Y.~N.}\ \bibnamefont
  {Ovchinnikov}},\ }\href@noop {} {\bibfield  {journal} {\bibinfo  {journal}
  {Sov. Phys. JETP}\ }\textbf {\bibinfo {volume} {28}},\ \bibinfo {pages}
  {1200} (\bibinfo {year} {1969})}\BibitemShut {NoStop}%
\bibitem [{\citenamefont {Serene}\ and\ \citenamefont
  {Rainer}(1983)}]{serene83}%
  \BibitemOpen
  \bibfield  {author} {\bibinfo {author} {\bibfnamefont {J.}~\bibnamefont
  {Serene}}\ and\ \bibinfo {author} {\bibfnamefont {D.}~\bibnamefont
  {Rainer}},\ }\href@noop {} {\bibfield  {journal} {\bibinfo  {journal}
  {Physics Reports}\ }\textbf {\bibinfo {volume} {101}},\ \bibinfo {pages}
  {221} (\bibinfo {year} {1983})}\BibitemShut {NoStop}%
\bibitem [{\citenamefont {Volovik}(1993)}]{volovik93}%
  \BibitemOpen
  \bibfield  {author} {\bibinfo {author} {\bibfnamefont {G.~E.}\ \bibnamefont
  {Volovik}},\ }\href@noop {} {\bibfield  {journal} {\bibinfo  {journal} {JETP
  Lett.}\ }\textbf {\bibinfo {volume} {58}},\ \bibinfo {pages} {469} (\bibinfo
  {year} {1993})}\BibitemShut {NoStop}%
\bibitem [{\citenamefont {Moler}\ \emph {et~al.}(1997)\citenamefont {Moler},
  \citenamefont {Sisson}, \citenamefont {Urbach}, \citenamefont {Beasley},
  \citenamefont {Kapitulnik}, \citenamefont {Baar}, \citenamefont {Liang},\
  and\ \citenamefont {Hardy}}]{KAMoler:1997}%
  \BibitemOpen
  \bibfield  {author} {\bibinfo {author} {\bibfnamefont {K.~A.}\ \bibnamefont
  {Moler}}, \bibinfo {author} {\bibfnamefont {D.~L.}\ \bibnamefont {Sisson}},
  \bibinfo {author} {\bibfnamefont {J.~S.}\ \bibnamefont {Urbach}}, \bibinfo
  {author} {\bibfnamefont {M.~R.}\ \bibnamefont {Beasley}}, \bibinfo {author}
  {\bibfnamefont {A.}~\bibnamefont {Kapitulnik}}, \bibinfo {author}
  {\bibfnamefont {D.~J.}\ \bibnamefont {Baar}}, \bibinfo {author}
  {\bibfnamefont {R.}~\bibnamefont {Liang}}, \ and\ \bibinfo {author}
  {\bibfnamefont {W.~N.}\ \bibnamefont {Hardy}},\ }\href {\doibase
  10.1103/PhysRevB.55.3954} {\bibfield  {journal} {\bibinfo  {journal} {Phys.
  Rev. B}\ }\textbf {\bibinfo {volume} {55}},\ \bibinfo {pages} {3954}
  (\bibinfo {year} {1997})}\BibitemShut {NoStop}%
\bibitem [{\citenamefont {Wang}\ \emph {et~al.}(2001)\citenamefont {Wang},
  \citenamefont {Revaz}, \citenamefont {Erb},\ and\ \citenamefont
  {Junod}}]{YWang:2001}%
  \BibitemOpen
  \bibfield  {author} {\bibinfo {author} {\bibfnamefont {Y.}~\bibnamefont
  {Wang}}, \bibinfo {author} {\bibfnamefont {B.}~\bibnamefont {Revaz}},
  \bibinfo {author} {\bibfnamefont {A.}~\bibnamefont {Erb}}, \ and\ \bibinfo
  {author} {\bibfnamefont {A.}~\bibnamefont {Junod}},\ }\href {\doibase
  10.1103/PhysRevB.63.094508} {\bibfield  {journal} {\bibinfo  {journal} {Phys.
  Rev. B}\ }\textbf {\bibinfo {volume} {63}},\ \bibinfo {pages} {094508}
  (\bibinfo {year} {2001})}\BibitemShut {NoStop}%
\bibitem [{\citenamefont {Ichioka}\ \emph {et~al.}(1996)\citenamefont
  {Ichioka}, \citenamefont {Hayashi}, \citenamefont {Enomoto},\ and\
  \citenamefont {Machida}}]{Ichioka96}%
  \BibitemOpen
  \bibfield  {author} {\bibinfo {author} {\bibfnamefont {M.}~\bibnamefont
  {Ichioka}}, \bibinfo {author} {\bibfnamefont {N.}~\bibnamefont {Hayashi}},
  \bibinfo {author} {\bibfnamefont {N.}~\bibnamefont {Enomoto}}, \ and\
  \bibinfo {author} {\bibfnamefont {K.}~\bibnamefont {Machida}},\ }\href
  {\doibase 10.1103/PhysRevB.53.15316} {\bibfield  {journal} {\bibinfo
  {journal} {Phys. Rev. B}\ }\textbf {\bibinfo {volume} {53}},\ \bibinfo
  {pages} {15316} (\bibinfo {year} {1996})}\BibitemShut {NoStop}%
\bibitem [{\citenamefont {Schopohl}\ and\ \citenamefont
  {Maki}(1995)}]{Schopohl95}%
  \BibitemOpen
  \bibfield  {author} {\bibinfo {author} {\bibfnamefont {N.}~\bibnamefont
  {Schopohl}}\ and\ \bibinfo {author} {\bibfnamefont {K.}~\bibnamefont
  {Maki}},\ }\href {\doibase 10.1103/PhysRevB.52.490} {\bibfield  {journal}
  {\bibinfo  {journal} {Phys. Rev. B}\ }\textbf {\bibinfo {volume} {52}},\
  \bibinfo {pages} {490} (\bibinfo {year} {1995})}\BibitemShut {NoStop}%
\bibitem [{\citenamefont {Ichioka}\ \emph {et~al.}(1997)\citenamefont
  {Ichioka}, \citenamefont {Hayashi},\ and\ \citenamefont
  {Machida}}]{Ichioka97}%
  \BibitemOpen
  \bibfield  {author} {\bibinfo {author} {\bibfnamefont {M.}~\bibnamefont
  {Ichioka}}, \bibinfo {author} {\bibfnamefont {N.}~\bibnamefont {Hayashi}}, \
  and\ \bibinfo {author} {\bibfnamefont {K.}~\bibnamefont {Machida}},\ }\href
  {\doibase 10.1103/PhysRevB.55.6565} {\bibfield  {journal} {\bibinfo
  {journal} {Phys. Rev. B}\ }\textbf {\bibinfo {volume} {55}},\ \bibinfo
  {pages} {6565} (\bibinfo {year} {1997})}\BibitemShut {NoStop}%
\bibitem [{\citenamefont {Ichioka}\ \emph {et~al.}(1999)\citenamefont
  {Ichioka}, \citenamefont {Hasegawa},\ and\ \citenamefont
  {Machida}}]{Ichioka99}%
  \BibitemOpen
  \bibfield  {author} {\bibinfo {author} {\bibfnamefont {M.}~\bibnamefont
  {Ichioka}}, \bibinfo {author} {\bibfnamefont {A.}~\bibnamefont {Hasegawa}}, \
  and\ \bibinfo {author} {\bibfnamefont {K.}~\bibnamefont {Machida}},\ }\href
  {\doibase 10.1103/PhysRevB.59.184} {\bibfield  {journal} {\bibinfo  {journal}
  {Phys. Rev. B}\ }\textbf {\bibinfo {volume} {59}},\ \bibinfo {pages} {184}
  (\bibinfo {year} {1999})}\BibitemShut {NoStop}%
\bibitem [{\citenamefont {Franz}\ and\ \citenamefont {Te\ifmmode \check{s}\else
  \v{s}\fi{}anovi\ifmmode~\acute{c}\else \'{c}\fi{}}(2000)}]{FranzTesanovic}%
  \BibitemOpen
  \bibfield  {author} {\bibinfo {author} {\bibfnamefont {M.}~\bibnamefont
  {Franz}}\ and\ \bibinfo {author} {\bibfnamefont {Z.}~\bibnamefont {Te\ifmmode
  \check{s}\else \v{s}\fi{}anovi\ifmmode~\acute{c}\else \'{c}\fi{}}},\ }\href
  {\doibase 10.1103/PhysRevLett.84.554} {\bibfield  {journal} {\bibinfo
  {journal} {Phys. Rev. Lett.}\ }\textbf {\bibinfo {volume} {84}},\ \bibinfo
  {pages} {554} (\bibinfo {year} {2000})}\BibitemShut {NoStop}%
\bibitem [{\citenamefont {Vekhter}\ and\ \citenamefont
  {Vorontsov}(2008)}]{Vekhter08}%
  \BibitemOpen
  \bibfield  {author} {\bibinfo {author} {\bibfnamefont {I.}~\bibnamefont
  {Vekhter}}\ and\ \bibinfo {author} {\bibfnamefont {A.}~\bibnamefont
  {Vorontsov}},\ }\href@noop {} {\bibfield  {journal} {\bibinfo  {journal}
  {Physica B: Condensed Matter}\ }\textbf {\bibinfo {volume} {403}},\ \bibinfo
  {pages} {958} (\bibinfo {year} {2008})}\BibitemShut {NoStop}%
\bibitem [{\citenamefont {Dahm}\ \emph {et~al.}(2002)\citenamefont {Dahm},
  \citenamefont {Graser}, \citenamefont {Iniotakis},\ and\ \citenamefont
  {Schopohl}}]{Dahm02}%
  \BibitemOpen
  \bibfield  {author} {\bibinfo {author} {\bibfnamefont {T.}~\bibnamefont
  {Dahm}}, \bibinfo {author} {\bibfnamefont {S.}~\bibnamefont {Graser}},
  \bibinfo {author} {\bibfnamefont {C.}~\bibnamefont {Iniotakis}}, \ and\
  \bibinfo {author} {\bibfnamefont {N.}~\bibnamefont {Schopohl}},\ }\href
  {\doibase 10.1103/PhysRevB.66.144515} {\bibfield  {journal} {\bibinfo
  {journal} {Phys. Rev. B}\ }\textbf {\bibinfo {volume} {66}},\ \bibinfo
  {pages} {144515} (\bibinfo {year} {2002})}\BibitemShut {NoStop}%
\bibitem [{\citenamefont {Schopohl}(1998)}]{schopohl98}%
  \BibitemOpen
  \bibfield  {author} {\bibinfo {author} {\bibfnamefont {N.}~\bibnamefont
  {Schopohl}},\ }\href@noop {} {\bibfield  {journal} {\bibinfo  {journal}
  {cond-mat/9804064 (unpublished)}\ } (\bibinfo {year} {1998})}\BibitemShut
  {NoStop}%
\bibitem [{Note2()}]{Note2}%
  \BibitemOpen
  \bibinfo {note} {Note that our notation of $g$, $f$, and $\protect
  \mathaccentV {bar}016{f}$ differs from the one used in Ref.~\protect
  \rev@citealpnum {schopohl98}. Under the transformation $g\to -i\pi g$, $f\to
  \pi f$, and $\protect \mathaccentV {bar}016{f}\to -\pi \protect \mathaccentV
  {bar}016{f}$ the notation in Ref.~\protect \rev@citealpnum {schopohl98}
  passes into our notation.}\BibitemShut {Stop}%
\bibitem [{\citenamefont {Suzuki}\ \emph {et~al.}(2011)\citenamefont {Suzuki},
  \citenamefont {Usui},\ and\ \citenamefont {Kuroki}}]{Kuroki2011}%
  \BibitemOpen
  \bibfield  {author} {\bibinfo {author} {\bibfnamefont {K.}~\bibnamefont
  {Suzuki}}, \bibinfo {author} {\bibfnamefont {H.}~\bibnamefont {Usui}}, \ and\
  \bibinfo {author} {\bibfnamefont {K.}~\bibnamefont {Kuroki}},\ }\href
  {\doibase 10.1143/JPSJ.80.013710} {\bibfield  {journal} {\bibinfo  {journal}
  {Journal of the Physical Society of Japan}\ }\textbf {\bibinfo {volume}
  {80}},\ \bibinfo {pages} {013710} (\bibinfo {year} {2011})}\BibitemShut
  {NoStop}%
\bibitem [{\citenamefont {Shimojima}\ \emph {et~al.}(2011)\citenamefont
  {Shimojima}, \citenamefont {Sakaguchi}, \citenamefont {Ishizaka},
  \citenamefont {Ishida}, \citenamefont {Kiss}, \citenamefont {Okawa},
  \citenamefont {Togashi}, \citenamefont {Chen}, \citenamefont {Watanabe},
  \citenamefont {Arita} \emph {et~al.}}]{Shimojima11}%
  \BibitemOpen
  \bibfield  {author} {\bibinfo {author} {\bibfnamefont {T.}~\bibnamefont
  {Shimojima}}, \bibinfo {author} {\bibfnamefont {F.}~\bibnamefont
  {Sakaguchi}}, \bibinfo {author} {\bibfnamefont {K.}~\bibnamefont {Ishizaka}},
  \bibinfo {author} {\bibfnamefont {Y.}~\bibnamefont {Ishida}}, \bibinfo
  {author} {\bibfnamefont {T.}~\bibnamefont {Kiss}}, \bibinfo {author}
  {\bibfnamefont {M.}~\bibnamefont {Okawa}}, \bibinfo {author} {\bibfnamefont
  {T.}~\bibnamefont {Togashi}}, \bibinfo {author} {\bibfnamefont
  {C.}~\bibnamefont {Chen}}, \bibinfo {author} {\bibfnamefont {S.}~\bibnamefont
  {Watanabe}}, \bibinfo {author} {\bibfnamefont {M.}~\bibnamefont {Arita}},
  \emph {et~al.},\ }\href@noop {} {\bibfield  {journal} {\bibinfo  {journal}
  {Science}\ }\textbf {\bibinfo {volume} {332}},\ \bibinfo {pages} {564}
  (\bibinfo {year} {2011})}\BibitemShut {NoStop}%
\bibitem [{\citenamefont {Mishra}\ \emph {et~al.}(2009)\citenamefont {Mishra},
  \citenamefont {Boyd}, \citenamefont {Graser}, \citenamefont {Maier},
  \citenamefont {Hirschfeld},\ and\ \citenamefont {Scalapino}}]{Mishra09}%
  \BibitemOpen
  \bibfield  {author} {\bibinfo {author} {\bibfnamefont {V.}~\bibnamefont
  {Mishra}}, \bibinfo {author} {\bibfnamefont {G.}~\bibnamefont {Boyd}},
  \bibinfo {author} {\bibfnamefont {S.}~\bibnamefont {Graser}}, \bibinfo
  {author} {\bibfnamefont {T.}~\bibnamefont {Maier}}, \bibinfo {author}
  {\bibfnamefont {P.~J.}\ \bibnamefont {Hirschfeld}}, \ and\ \bibinfo {author}
  {\bibfnamefont {D.~J.}\ \bibnamefont {Scalapino}},\ }\href {\doibase
  10.1103/PhysRevB.79.094512} {\bibfield  {journal} {\bibinfo  {journal} {Phys.
  Rev. B}\ }\textbf {\bibinfo {volume} {79}},\ \bibinfo {pages} {094512}
  (\bibinfo {year} {2009})}\BibitemShut {NoStop}%
\bibitem [{\citenamefont {{Mishra}}\ \emph {et~al.}(2011)\citenamefont
  {{Mishra}}, \citenamefont {{Graser}},\ and\ \citenamefont
  {{Hirschfeld}}}]{Mishra2011}%
  \BibitemOpen
  \bibfield  {author} {\bibinfo {author} {\bibfnamefont {V.}~\bibnamefont
  {{Mishra}}}, \bibinfo {author} {\bibfnamefont {S.}~\bibnamefont {{Graser}}},
  \ and\ \bibinfo {author} {\bibfnamefont {P.~J.}\ \bibnamefont
  {{Hirschfeld}}},\ }\href@noop {} {\bibfield  {journal} {\bibinfo  {journal}
  {ArXiv e-prints}\ } (\bibinfo {year} {2011})},\ \Eprint
  {http://arxiv.org/abs/1101.5699} {arXiv:1101.5699 [cond-mat.supr-con]}
  \BibitemShut {NoStop}%
\bibitem [{\citenamefont {Yoshida}\ \emph {et~al.}(2011)\citenamefont
  {Yoshida}, \citenamefont {Nishi}, \citenamefont {Ideta}, \citenamefont
  {Fujimori}, \citenamefont {Kubota}, \citenamefont {Ono}, \citenamefont
  {Kasahara}, \citenamefont {Shibauchi}, \citenamefont {Terashima},
  \citenamefont {Matsuda}, \citenamefont {Ikeda},\ and\ \citenamefont
  {Arita}}]{Yoshida11}%
  \BibitemOpen
  \bibfield  {author} {\bibinfo {author} {\bibfnamefont {T.}~\bibnamefont
  {Yoshida}}, \bibinfo {author} {\bibfnamefont {I.}~\bibnamefont {Nishi}},
  \bibinfo {author} {\bibfnamefont {S.}~\bibnamefont {Ideta}}, \bibinfo
  {author} {\bibfnamefont {A.}~\bibnamefont {Fujimori}}, \bibinfo {author}
  {\bibfnamefont {M.}~\bibnamefont {Kubota}}, \bibinfo {author} {\bibfnamefont
  {K.}~\bibnamefont {Ono}}, \bibinfo {author} {\bibfnamefont {S.}~\bibnamefont
  {Kasahara}}, \bibinfo {author} {\bibfnamefont {T.}~\bibnamefont {Shibauchi}},
  \bibinfo {author} {\bibfnamefont {T.}~\bibnamefont {Terashima}}, \bibinfo
  {author} {\bibfnamefont {Y.}~\bibnamefont {Matsuda}}, \bibinfo {author}
  {\bibfnamefont {H.}~\bibnamefont {Ikeda}}, \ and\ \bibinfo {author}
  {\bibfnamefont {R.}~\bibnamefont {Arita}},\ }\href {\doibase
  10.1103/PhysRevLett.106.117001} {\bibfield  {journal} {\bibinfo  {journal}
  {Phys. Rev. Lett.}\ }\textbf {\bibinfo {volume} {106}},\ \bibinfo {pages}
  {117001} (\bibinfo {year} {2011})}\BibitemShut {NoStop}%
\bibitem [{\citenamefont {Kramer}\ and\ \citenamefont
  {Pesch}(1974)}]{KramerPesch1}%
  \BibitemOpen
  \bibfield  {author} {\bibinfo {author} {\bibfnamefont {L.}~\bibnamefont
  {Kramer}}\ and\ \bibinfo {author} {\bibfnamefont {W.}~\bibnamefont {Pesch}},\
  }\href@noop {} {\bibfield  {journal} {\bibinfo  {journal} {Z. Phys.}\
  }\textbf {\bibinfo {volume} {269}},\ \bibinfo {pages} {59} (\bibinfo {year}
  {1974})}\BibitemShut {NoStop}%
\bibitem [{\citenamefont {Pesch}\ and\ \citenamefont
  {Kramer}(1974)}]{KramerPesch2}%
  \BibitemOpen
  \bibfield  {author} {\bibinfo {author} {\bibfnamefont {W.}~\bibnamefont
  {Pesch}}\ and\ \bibinfo {author} {\bibfnamefont {L.}~\bibnamefont {Kramer}},\
  }\href@noop {} {\bibfield  {journal} {\bibinfo  {journal} {J. Low T. Phys.}\
  }\textbf {\bibinfo {volume} {15}},\ \bibinfo {pages} {367} (\bibinfo {year}
  {1974})}\BibitemShut {NoStop}%
\bibitem [{\citenamefont {Pesch}(1975)}]{Pesch_approx1}%
  \BibitemOpen
  \bibfield  {author} {\bibinfo {author} {\bibfnamefont {W.}~\bibnamefont
  {Pesch}},\ }\href@noop {} {\bibfield  {journal} {\bibinfo  {journal} {Z.
  Phys. B}\ }\textbf {\bibinfo {volume} {21}},\ \bibinfo {pages} {263}
  (\bibinfo {year} {1975})}\BibitemShut {NoStop}%
\bibitem [{\citenamefont {Zhitomirsky}\ and\ \citenamefont
  {Dao}(2004)}]{Zhitomirsky:2004}%
  \BibitemOpen
  \bibfield  {author} {\bibinfo {author} {\bibfnamefont {M.~E.}\ \bibnamefont
  {Zhitomirsky}}\ and\ \bibinfo {author} {\bibfnamefont {V.-H.}\ \bibnamefont
  {Dao}},\ }\href {\doibase 10.1103/PhysRevB.69.054508} {\bibfield  {journal}
  {\bibinfo  {journal} {Phys. Rev. B}\ }\textbf {\bibinfo {volume} {69}},\
  \bibinfo {pages} {054508} (\bibinfo {year} {2004})}\BibitemShut {NoStop}%
\bibitem [{\citenamefont {Klimesch}\ and\ \citenamefont
  {Pesch}(1978)}]{Pesch_approx2}%
  \BibitemOpen
  \bibfield  {author} {\bibinfo {author} {\bibfnamefont {P.}~\bibnamefont
  {Klimesch}}\ and\ \bibinfo {author} {\bibfnamefont {W.}~\bibnamefont
  {Pesch}},\ }\href@noop {} {\bibfield  {journal} {\bibinfo  {journal} {J. Low
  T. Phys.}\ }\textbf {\bibinfo {volume} {32}},\ \bibinfo {pages} {869}
  (\bibinfo {year} {1978})}\BibitemShut {NoStop}%
\end{thebibliography}
\end{document}